\documentclass[useAMS,usenatbib]{mn2e}

\input{epsf}

\usepackage{natbib}
\usepackage{graphicx}	
\usepackage{amsmath}
\usepackage{longtable}
\usepackage{lscape}
\usepackage{threeparttable}
\usepackage{threeparttablex}
\usepackage{deluxetable}
\usepackage[singlelinecheck=false]{caption}
\usepackage{afterpage}

\newcommand{\CIVw}{C\,{\sc iv}~$\lambda$1549}

\newcommand{\OIIIw}{[O$\,\textsc{iii}]$~$\lambda$5007}

\newcommand{\CIV}{C\,{\sc iv}}

\newcommand{\Hb}{{H}\,$\beta$}
\newcommand{\FeII}{Fe\,\textsc{ii}}

\newcommand{\OIII}{[O\,{\sc iii}]}

\newcommand{\siivoiv}{Si\,{\sc iv}+O\,{\sc iv}]}

\newcommand{\fwhmhb}{$\rm{FWHM}_{\rm{H}\beta}$}
\newcommand{\fwhmhbpred}{$\rm{FWHM}_{\rm{H}\beta,\rm{predicted}}$}

\newcommand{\fwhmciv}{FWHM$_{\rm{C\,{\sc IV}}}$}

\def\lsim{\lower0.3em\hbox{$\,\buildrel <\over\sim\,$}}
\def\gsim{\lower0.3em\hbox{$\,\buildrel >\over\sim\,$}}

\title[\CIV\ and orientation in quasars]{The behavior of quasar \CIV\ emission-line properties with orientation}
\author[J. C. Runnoe et al.]{Jessie C. Runnoe$^{1}$\thanks{E-mail:
jrunnoe@uwyo.edu} , M. S. Brotherton$^{1}$, M. A. DiPompeo$^1$, and Zhaohui Shang$^{2}$\\
$^{1}$Department of Physics and Astronomy, University of Wyoming, Laramie, WY 82071, USA\\
$^{2}$Department of Physics, Tianjin Normal University, Tianjin 300387, China}
\begin{document}		

\date{Preprint 2012 October 13}

\pagerange{\pageref{firstpage}--\pageref{lastpage}} \pubyear{2012}

\maketitle

\label{firstpage}

\begin{abstract}
With a quasar sample designed for studying orientation effects, we investigate the orientation dependence of characteristics of the \CIVw\ broad emission line in approximately 50 Type 1 quasars with $z=0.1-1.4$.  Understanding the role that orientation plays in determining the observed ultraviolet spectra in quasars is of interest both for the insight it gives into the physical emitting regions and for its ramifications for the estimation of fundamental properties.  Orientation is measured for the sample via radio core dominance.  In our analysis we include measurements of the continuum luminosity and the optical-to-X-ray spectral slope, spectral properties commonly included in the suite known as ``Eigenvector 1'', and the full-width at half maximum, full-width at one-quarter-maximum, shape, blueshift, and equivalent width of the \CIV\ broad emission line.  We also investigate a new prescription that we recently developed for predicting the velocity line width of the \Hb\ broad emission line based on the velocity line width of the \CIV\ line and the ratio of continuum subtracted peak fluxes of \siivoiv\ at 1400 \AA\ to \CIV.  

We find that, while it is known that orientation does not drive Eigenvector 1, it does affect some properties included in Eigenvector 1.  Specifically, we find marginal orientation dependencies for the equivalent width of \CIV and the equivalent width ratio of the optical \FeII\ to \OIIIw.  Although, in the case of the \CIV\ equivalent width, this may actually be due to an orientation dependence of the ultraviolet continuum luminosity which is found to be marginally significant.  The full-width at one-quarter-maximum also shows a marginally significant orientation dependence.  Notably, the shape and blueshift of \CIV\ and $\alpha_{ox}$ do not depend on orientation.  In addition to a correlation analysis of the ultraviolet spectral properties and radio core dominance, we provide composite spectra of edge-on and face-on sources for this sample.  In particular, we highlight the orientation dependence of the velocity line width predicted for \Hb.  We find that this predicted line width depends on orientation in a manner similar to the true velocity line width of \Hb, where no such dependence is observed for \CIV.  This is an indication that orientation information concerning the line emitting regions can be extracted from ultraviolet spectra.  
\end{abstract}


\section{introduction}
In the current quasar paradigm,  an active galactic nucleus (AGN) is powered by a central supermassive black hole.  Surrounding the black hole is an accretion disk \citep{urry95} that acts as a continuum source and photoionizes high-velocity gas in the broad line region (BLR) to produce the characteristic broad emission lines that are observed in the spectra of Type 1 AGN.  These central structures are not resolved in images, so information about the velocity field, ionization, and geometry of the BLR must be gleaned from other techniques.  Despite a rich literature on the subject \citep[e.g.,][]{osterbrock86,collin88,baldwin95,peterson99,sulentic00a} the BLR is still poorly understood. 

Reverberation mapping, the technique of observing time lags between variation in the continuum emission and broad emission lines, has been a fruitful source of results regarding the structure of the BLR \citep{peterson93,netzer97}.  Confined to relatively small radii, the BLR is geometrically thick (where the inner radius of the BLR is much smaller than the outer radius) and has a stratified ionization structure.  Shorter time lags are observed for high-ionization lines like \CIVw, indicating that this line emitting gas is located closer to the central engine than the gas that emits low-ionization lines like \Hb.  Reverberation mapping also suggests that the BLR is virialized, with the gravitational potential of the supermassive black hole dominating the motions of the line emitting gas \citep[e.g.,][]{peterson99}.  As a result, the characteristic velocity widths of lines like \CIV\ are expected to be broader than lower-ionization lines like \Hb, but still tightly correlated \citep{peterson11}.  In single-epoch spectra this is not observed, and in the case of \CIV\ and \Hb\ there is significant scatter between velocity widths measured from the two lines and disagreement from the stratification picture with \CIV\ sometimes displaying widths significantly narrower than \Hb\ \citep[e.g.,][]{shang07}.  This discrepancy exposes the limits of our understanding of the BLR and hints that the nature of the \CIV\ emitting gas is not simple.

Statistical investigations of broad emission line profiles further reveal the complexity of the BLR kinematics and geometry, and in particular differences in the configuration of the \CIV\ and \Hb\ emitting gas.  Compared to \Hb, the line profile of \CIV\ can be asymmetric and is often blueshifted from the systematic redshift of the object \citep[e.g.,][]{gaskell82,wilkes82,tytler92,richards02}.  The \CIV-emitting gas is unlikely to be found in a pure disk-like configuration.  Besides being unable to produce the observed blueshifts and line asymmetries, \citet{fine10} rule out a pure disk based on the small dispersion found in the distribution of \CIV\ line widths.  By binning objects by the full-width at half-maximum (FWHM) of \CIV, \citet{wills93} illustrate the variation in the shape of the \CIV\ line between broad, boxy and narrower, peaky profiles.  \citet{denney12} found that, while the boxy \CIV\ profiles are primarily composed of emission that reverberates in response to variation in the continuum source, the peaky profiles contain a significant contribution from a low-velocity, non-reverberating core contaminant.  In contrast, the complete \Hb\ line profile reverberates in response to continuum variation with the exception of a more easily characterized contribution from the narrow line region (NLR).  This evidence suggests that, unlike \Hb, the \CIV\ line profile may not be determined exclusively or even primarily by virialized gas in the BLR and has a contaminating component that is often significant \citep[e.g.,][]{denney12}. 

The incomplete understanding of the \CIV\ emitting gas is propagated into black hole masses estimated for thousands of objects via the single-epoch black hole mass scaling relationships \citep[e.g.,][]{vestergaard06,park13}.  The reverberation mapping-based scaling relationships are calibrated for multiple emission lines, \Hb\ and \CIV\ among others, and take a characteristic velocity and nearby continuum luminosity in order to estimate the mass of the central supermassive black hole.  The \Hb\ line has the best calibration, but with increasing redshift, \Hb\ moves out of the optical window and \CIV\ must be used instead.  Thus, in order to probe the so-called peak of quasar activity around $z\sim2$, reliable \CIV-based black hole masses are required.  The disagreement between characteristic velocities calculated from \Hb\ and \CIV\ casts doubt on the reliability of the \CIV-based black hole mass estimates \citep[e.g,][]{trakhtenbrot12}.  The literature suggests that this may be due largely to the fact that the \CIV-emitting structures are complex and the treatment used to extract BLR velocities from the \Hb\ line is not appropriate for use on \CIV\ \citep[e.g.,][]{denney12,wills93}.  

In order to improve agreement between velocity line widths calculated from \Hb\ and \CIV, \citet[][hereafter Paper I]{runnoe13c} recently developed a new methodology for predicting the characteristic \Hb\ velocity based on the spectral analysis of \citet{wills93}.  It has been noted that the strength of the blend of \siivoiv\ at 1400 \AA\ (hereafter $\lambda$1400) is relatively constant even as the \CIV\ line changes shape and strength \citep[e.g.,][]{wills93,richards02}.  The presence of contamination in the \CIV\ line largely acts to bring up the line peak.  Thus the ratio of continuum subtracted peak fluxes, Peak($\lambda$1400/\CIV), indicates the strength of the low-velocity, non-reverberating gas and the amount of contamination in the \CIV\ line.  The prescription is given by

{\scriptsize
\begin{eqnarray}
\label{eqn:EV1}
\nonumber {\rm log}\left[\frac{\textrm{\fwhmhbpred}}{\textrm{km s}^{-1}}\right] &=& \rm{log}\left[\frac{\textrm{\fwhmciv}}{\textrm{km s}^{-1}}\right] - (   0.366\pm   0.048) \\
&-& (   0.574\pm   0.061)\,\rm{log}\left[\textrm{Peak}\left(\frac{\lambda1400}{\textrm{\CIV}}\right)\right],
\end{eqnarray}}

\noindent and produces line widths that show better agreement than the FWHM of \CIV\ with the observed \Hb\ FWHM.  This new methodology accounts for the fact that, though the line width of \Hb\ depends primarily on the black hole mass, the line width of \CIV\ also has a strong contaminant that scales with Peak($\lambda$1400/\CIV).  

This peak flux ratio is known to be among a suite of spectral parameters collectively known as ``Eigenvector 1'' (EV1) that efficiently describes the largest variation between quasar spectra.  Originally expressed in the optical by \citet{bg92}, EV1 is dominated by the anti-correlation of optical \FeII\ and \OIIIw\ emission but has ultraviolet (UV) indicators including Peak($\lambda$1400/\CIV) \citep{brotherton99a,shang03,sulentic07}.  The physical driver of EV1 remains elusive, although black hole mass is an unlikely candidate as objects with similar black hole masses exhibit a range of EV1 properties \citep{boroson02}.  If EV1 is primarily driven by Eddington fraction ($L/L_{Edd}$), as \citet{boroson02} suggest, certain sample selections can induce a correlation between EV1 and black hole mass that is not physically real.

One way of gaining significant insight into the \Hb\ and \CIV\ emitting regions is to investigate the orientation dependencies of these emission lines, with the velocity line widths being of particular interest.  The FWHM of \Hb\ depends on orientation, as measured by radio core dominance, in a way that is consistent with the \Hb-emitting gas being configured in an axisymmetric disk \citep{wills86}.  In previous studies, no orientation dependence in the FWHM of \CIV\ has been observed \citep{vestergaard00,runnoe13a}, but \citet{vestergaard00} do find a significant correlation for the full-width at twenty-percent-maximum (FW20M), similar to the full-width at one-quarter-maximum (FWQM), with radio-based orientation indicators.  It may be that the FWQM measurement targets the emission at the base of the \CIV\ line from virialized gas and hints at an orientation dependence similar to what is observed for \Hb, but the FWHM measurement is significantly contaminated by the presence of the non-virialized gas which makes it difficult to tease out. The FWHM predicted for \Hb\ via Equation~\ref{eqn:EV1} is expected to demonstrate the same orientation dependence as the observed FWHM of \Hb, therefore indicating that the prescription does a good job of emulating the observed \Hb\ behavior on multiple fronts such as velocity width and orientation dependence.  More subtly, this will also indicate that there is information about an orientation dependence in the BLR that is present in the UV part of the spectrum and can be teased out.

The aim of this work is to investigate the orientation dependence of the \CIV\ line profile with a particular focus on the characteristic \Hb\ velocity predicted via the Paper I prescription using the FWHM of CIV and Peak($\lambda$1400/\CIV).  The spectral energy distribution (SED) atlas of \citet{shang11} is particularly suited to this endeavor with its quasi-simultaneous optical/UV spectrophotometry and radio-loud (RL) subsample that was designed for studying orientation.  

This paper is structured as follows.  In Section~\ref{sec:data} we present the SED sample and relevant spectral measurements, in Section~\ref{sec:analysis} we detail our analysis which we discuss in the context of other work in Section~\ref{sec:discussion}, and in Section~\ref{sec:conclusion} we summarize this investigation.  Throughout this work we use a cosmology with $H_0 = 70$ km s$^{-1}$ Mpc$^{-1}$, $\Omega_{\Lambda} = 0.7$, and $\Omega_{m} = 0.3$.

\section{Sample, Data, and Measurements}
\label{sec:data}

We employ the RL subsample of the \citet{shang11} SED atlas for this investigation.  This sample was also used in \citet{runnoe13a} to investigate the orientation dependence of single-epoch black hole mass scaling relationships, so we refer the reader to that work for additional details and measurements.  This sample originates with an early {\it Hubble Space Telescope} ({\it HST}) program where objects were selected to have similar luminosity of the extended radio structures, a property thought to be isotropic, so that variation in the radio core luminosity is indicative of variation in the orientation of the source relative to the line of sight.  The blazars originally included in this sample, which are unsuitable for emission-line investigations due to rapid optical/UV variability from a beamed synchrotron jet, have been removed leaving a sample of 52 objects \citep{runnoe13a}.  This sample is ideal for investigating orientation in quasars because it contains objects similar in intrinsic power viewed at different angles.  We note that, because all sources are RL, this sample shows a limited range in EV1 properties which should be kept in mind throughout the analysis.  Additional details on this sample can be found in \citet{wills95} and \citet{netzer95}.

In order to estimate orientation in this sample, we use the radio core dominance parameter, log$\,R$, which is the ratio of the flat-spectrum core to the steep-spectrum lobes measured at rest-frame 5~GHz.  This parameter uses the amount of relativistic beaming in the core to estimate the angle between the line of sight and the radio jet, although there is significant scatter involved in calculating the angle from log$\,R$ \citep{ghisellini93}.  In general, positive values of log$\,R$ indicate a face-on source and negative values of log$\,R$ indicate a more edge-on source.  We have measured additional orientation indicators for this sample, but in \citet{runnoe13a} we did not find that they exhibited significantly different behavior so we do not include them here.  Our sample does not include blazars with very small inclination angles near $0^\circ$ or truly edge-on sources with inclinations near $90^\circ$ as these views are blocked by the dusty torus.

Optical spectrophotometry for this sample was collected quasi-simultaneously with the UV {\it HST} observations from either Kitt Peak National Observatory or McDonald Observatory, making this sample a powerful tool for comparing physical parameters measured from optical and UV emission lines.  The lag between the optical and UV data collection was on the order of weeks or less and fluxes matched to within a few percent.  X-ray coverage for the sample comes from either the {\it Chandra X-ray Observatory}, {\it XMM-Newton}, or {\it ROSAT} and is available for 34 out of 52 (65\%) sources in the RL subsample.  The time between optical/UV and X-ray observations varies, but is typically on the order of years.

We adopt the specific spectral decompositions performed and presented by \citet{tang12} and Paper I.  Further details of the fitting methods are outlined in  \citet{shang05} and \citet{shang07}.  Briefly, the wavelength windows in the vicinity of  $\lambda$1400, \CIV, and \Hb\ were fit separately with a power-law continuum and two generally broad Gaussians per broad emission line to reproduce the line profiles.  Note that this means the \CIV\ and $\lambda$1400 continua were fitted separately because each broad emission
line was decomposed independently.  For $\lambda$1400, where the line is blended \siivoiv, and \CIV, which is a doublet, this means that four Gaussians are used.  In these cases, the Gaussian pairs used to model each broad emission line are identical except in their centroids which are appropriate for the relevant line centers.  Note also that the \CIV\ and $\lambda$1400 regions were fit with separate continua.  The fit to \Hb\ includes an additional narrow Gaussian to model the \Hb\ emission from the NLR.  \citet{wills93} demonstrates that the NLR \CIV\ emission is weak so we do not include such a component in our fits, although some still model the NLR emission in \CIV\ \citep[e.g.,][]{sulentic07}.  In the \Hb\ region, we also include an \FeII\ template based on the narrow-line Seyfert 1 I Zw1 \citep{bg92}.  The optical \FeII\ template of \citet{veron-cetty04} has some differences from the \citet{bg92} template, particularly the inclusion of narrow lines around $\lambda \simeq 5000$ \AA\ and $\lambda \gsim 6400$ \AA.  As all of these sources are RL, they typically have weak \FeII\ emission minimizing the impact of selecting one template or the other.

The resulting fits are used to calculate both physical and spectral parameters for the sample.  EV1 indicators, including the  optical equivalent width (EW) ratio EW(\FeII/\OIII) and the UV ratio Peak($\lambda$1400/\CIV), are given in Paper I.  Peak($\lambda$1400/\CIV) is measured by taking the ratio of the continuum-subtracted peak fluxes of the line profiles of $\lambda$1400 and \CIV.  Also presented in that work is the prescription, listed here in Equation~\ref{eqn:EV1}, we use for predicting the velocity of the virialized \Hb\ BLR gas from the \CIV\ FWHM and Peak($\lambda$1400/\CIV).

The optical-to-X-ray slope, $\alpha_{ox}$, we calculate following the definition given in \citet{tananbaum79}.  We emphasize that this parameter is calculated only in sources with optical and X-ray coverage, so the calculation reflects the observed slope and does not rely on the fact that optical and X-ray luminosities are tightly correlated \citep[e.g.,][]{steffen06}.  The X-ray fluxes used in this calculation are tabulated by \citet{shang11} from the literature, so we refer the reader to that work for additional details of the X-ray analysis.  

Redshifts are measured in \citet{shang11} from the centroid of the \OIII\ narrow emission line.  As a check, the result is compared to the rest-frame established by other available narrow lines in each spectrum.  The accuracy obtained by this method is generally 0.0002 for most objects.

To these EV1 and line width measurements, we add other parameters potentially of interest in the context of \CIV\ that have already been measured for this sample.  In addition to a characteristic BLR velocity, estimates of black hole mass in single-epoch spectra require the monochromatic luminosity near 1450 \AA\ motivating us to include this parameter which was originally presented in \citet{runnoe12a}.  Furthermore, others have identified additional parameters that play a role in determining the observed \CIV\ line profile that are of interest when investigating the orientation dependence of \CIV.  We therefore also include three parameters integral to the disk-wind model of \citet{richards11}: the slope between the flux at 2500~\AA\ and 2~keV, $\alpha_{ox}$, the EW of \CIV, and the blueshift of \CIV.  In this model, $\alpha_{ox}$ estimates the relative strength of the X-ray luminosity compared to the optical/UV which determines the \CIV\ line profile via the relative strengths of the disk and wind components as indicated by the EW and blueshift of \CIV, respectively.  This occurs because an elevated X-ray luminosity may increase ionization levels and in turn dampen the radiative line driving that accelerates the wind, effectively weakening this component in the model.  EW and blueshift measurements were presented in \citet{tang12}, where the blueshift is measured relative to the systematic redshift of the source, and $\alpha_{ox}$ is measured directly from the data and presented for the first time in this work.  The shape of the \CIV\ line profile is a UV EV1 indicator that we include based on the analysis in \citet{denney12}.  The measurement of the line shape, $S=\textrm{FWHM}/\sigma_{l}$ where $\sigma_{l}$ is the line dispersion, was originally presented in Paper I.  Finally, we include the FWQM of \CIV\ to investigate the orientation dependence originally noted by \citet{vestergaard00}.  

We estimate uncertainties on spectral measurements and take care not to underestimate these values.  In high signal-to-noise spectra, formal uncertainties of the fitting process can be very small but can misleading as they may not dominate the error budget.  For example, the application of two different fitting methods to the same spectrum can yield measurements that are not consistent within the formal uncertainties \citep[e.g.,][fig. 4]{denney12}.  In order to avoid underestimating uncertainties on our spectral measurements we estimate the uncertainties rather than providing the formal uncertainties.  For fluxes we take the uncertainty in the flux calibration quoted by \citet{shang11} based on variability considerations and standard spectrophotometric reduction practices, for line widths we compare to published measurements for some representative objects, and for the blueshifts we appeal to the redshift uncertainty.  The uncertainties on calculated parameters are then propagated from these values.  The resulting values, listed below in the table notes, are meant to be conservative estimates of uncertainty that account for both formal and systematic sources of error.

All the spectral measurements for this sample are tabulated in Table~\ref{tab:measurements}, which is arranged as follows:
\begin{itemize}
\item[] Column (1) gives the object name.
\item[] Column (2) gives the EW ratio of \FeII\ to \OIII, a measure of EV1.  Typical measurement uncertainties are on the order of 40\%.
\item[] Column (3) gives the FWHM of \Hb.  Typical measurement uncertainties are on the order of 15\%.
\item[] Column (4) gives the shape of \CIV, $S=\textrm{FWHM}/\sigma_{l}$.  Typical measurement uncertainties are on the order of 20\%.   
\item[] Column (5) gives the gives the blueshift of \CIV\ measured relative to the systematic redshift of the source.  Typical measurement uncertainties are on the order of 100 km s$^{-1}$.
\item[] Columns (6) and (7) give the FWQM and FWHM of \CIV, respectively.  Typical measurement uncertainties are on the order of 15\%.
\item[] Column (8) gives the EW of \CIV.  Typical measurement uncertainties are on the order of 10\%.
\item[] Column (9) gives the FWHM predicted for \Hb\ from the FWHM of \CIV\ and the ratio Peak($\lambda$1400/\CIV).  Typical measurement uncertainties are on the order of 17\%.
\item[] Column (10) gives the ratio of the continuum subtracted peak flux of $\lambda$1400 to \CIV.  Typical measurement uncertainties are on the order of 15\%.
\item[] Column (11) gives $\alpha_{ox}$, the slope between the optical and X-ray regions of the SED.  Typical measurement uncertainties are on the order of 7\% without accounting for variability effects which could cause significantly increase the uncertainty in this parameter.
\item[] Column (12) gives the monochromatic luminosity at 1450 \AA.  Typical measurement uncertainties are on the order of 5\%.
\item[] Column (13) gives the radio core dominance, log$\,R$.  Typical measurement uncertainties are on the order of 15\%.
\end{itemize}



Given the sample selection and the measurements described above we are well poised to use an investigation of the orientation dependencies of UV spectral properties to gain insight into the \CIV\ emitting regions.

\section{Analysis}
\label{sec:analysis}
We investigated the orientation dependence of a variety of spectral parameters of, or commonly associated with, the \CIV\ line profile.  For this analysis we have employed two approaches, a rank correlation analysis and the construction of composite spectra.  The first gives insight into the relationships between the tabulated properties and the second visually displays spectral differences as a function of orientation.

\subsection{Correlation analysis}
\begin{figure*}
\begin{minipage}[!b]{7cm}
\centering
\includegraphics[width=7cm]{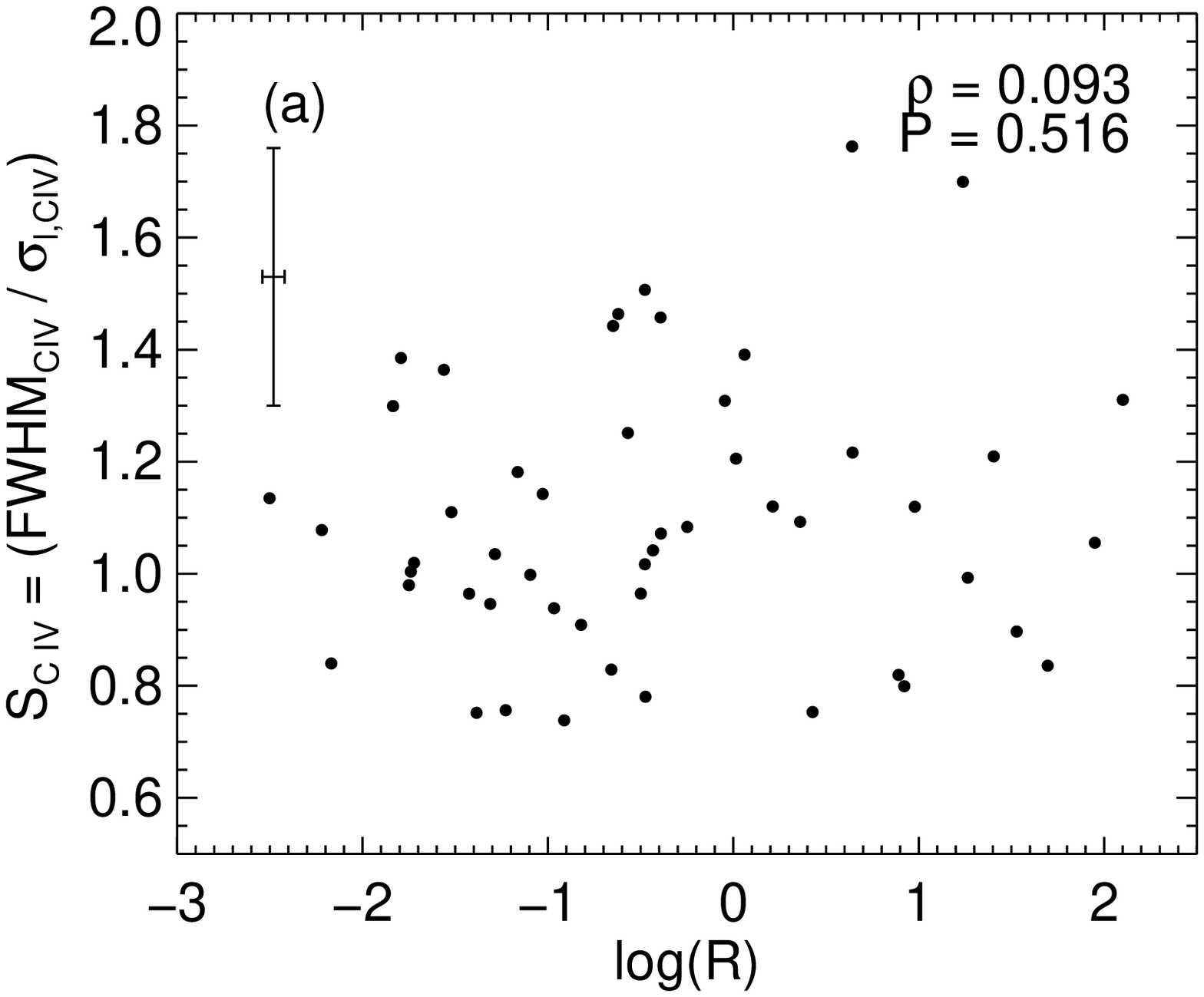}
\end{minipage}
\hspace{0.7cm}
\begin{minipage}[!b]{7cm}
\centering
\includegraphics[width=7cm]{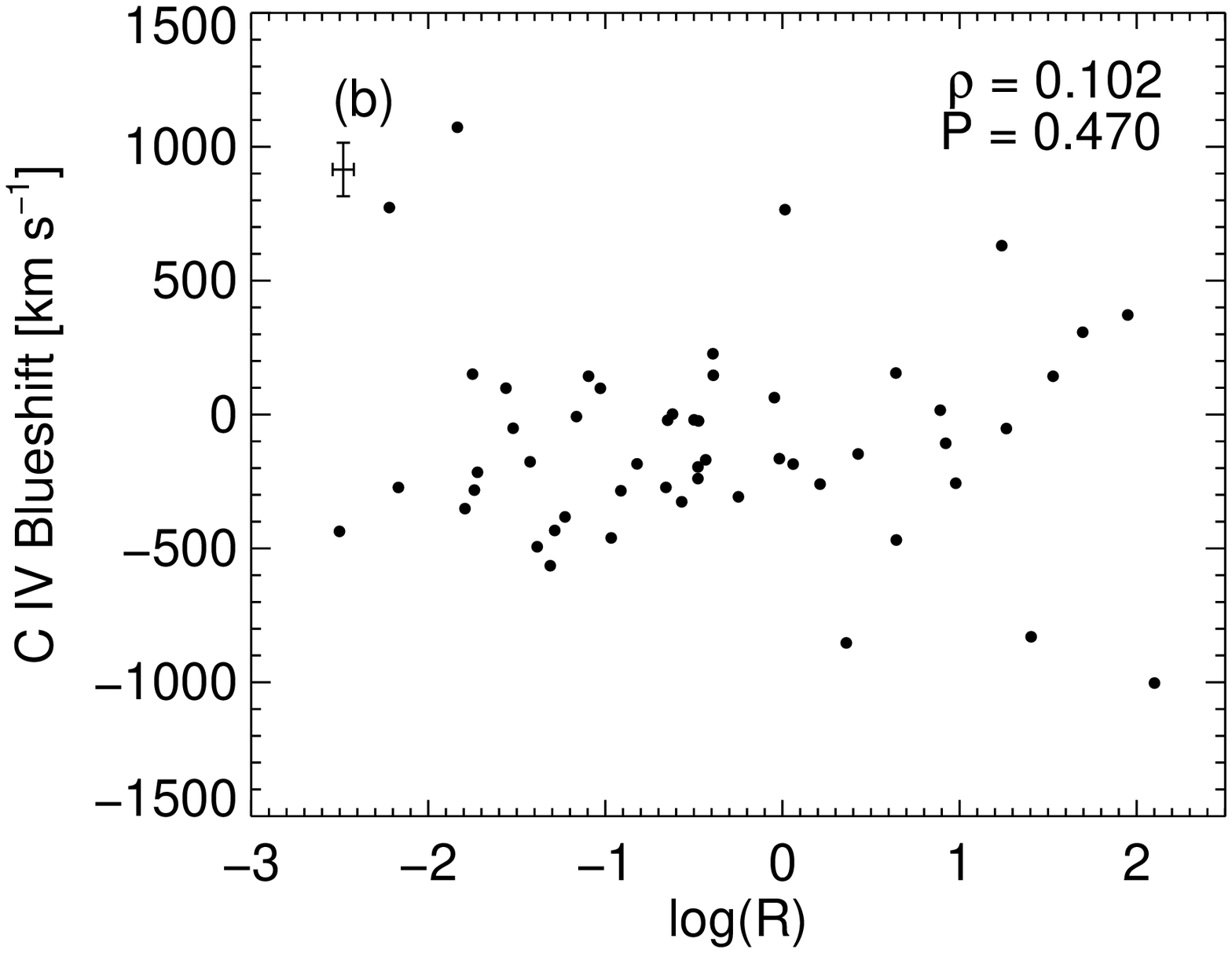}
\end{minipage}
\hspace{0.7cm}
\begin{minipage}[!b]{7cm}
\centering
\includegraphics[width=7cm]{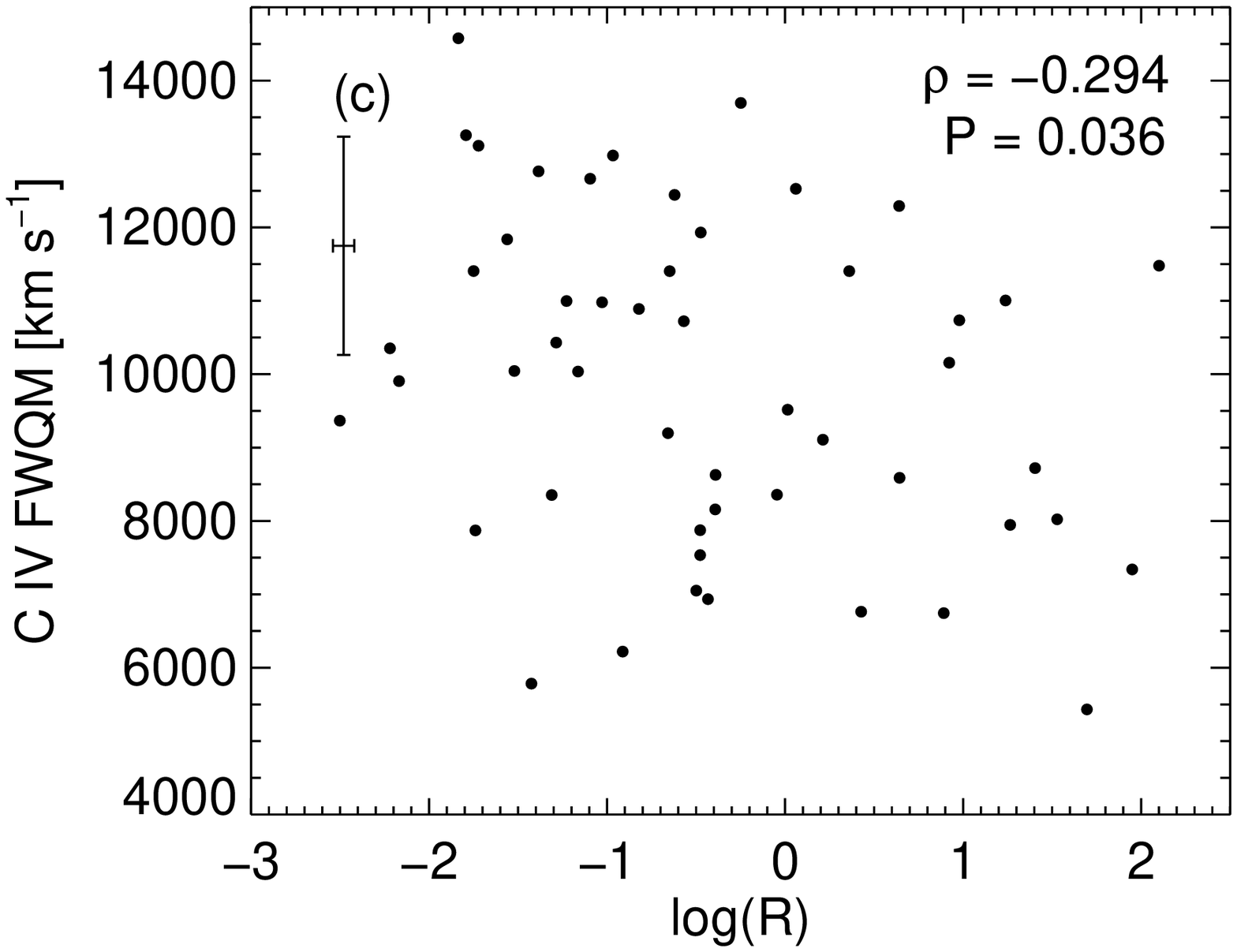}
\end{minipage}     
\hspace{0.7cm}       
\begin{minipage}[!b]{7cm}
\centering
\includegraphics[width=7cm]{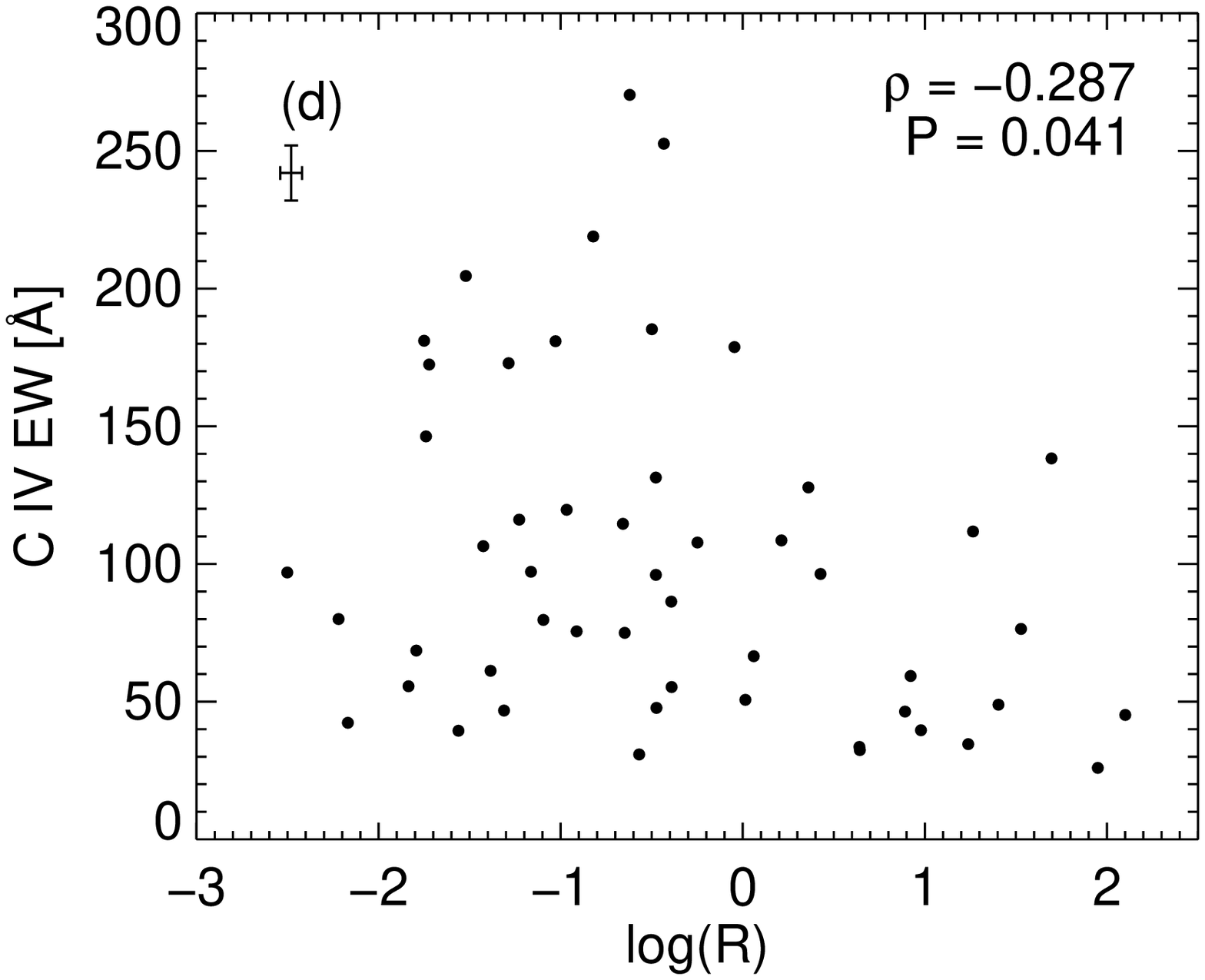}
\end{minipage}           
\begin{minipage}[!b]{7cm}
\centering
\includegraphics[width=7cm]{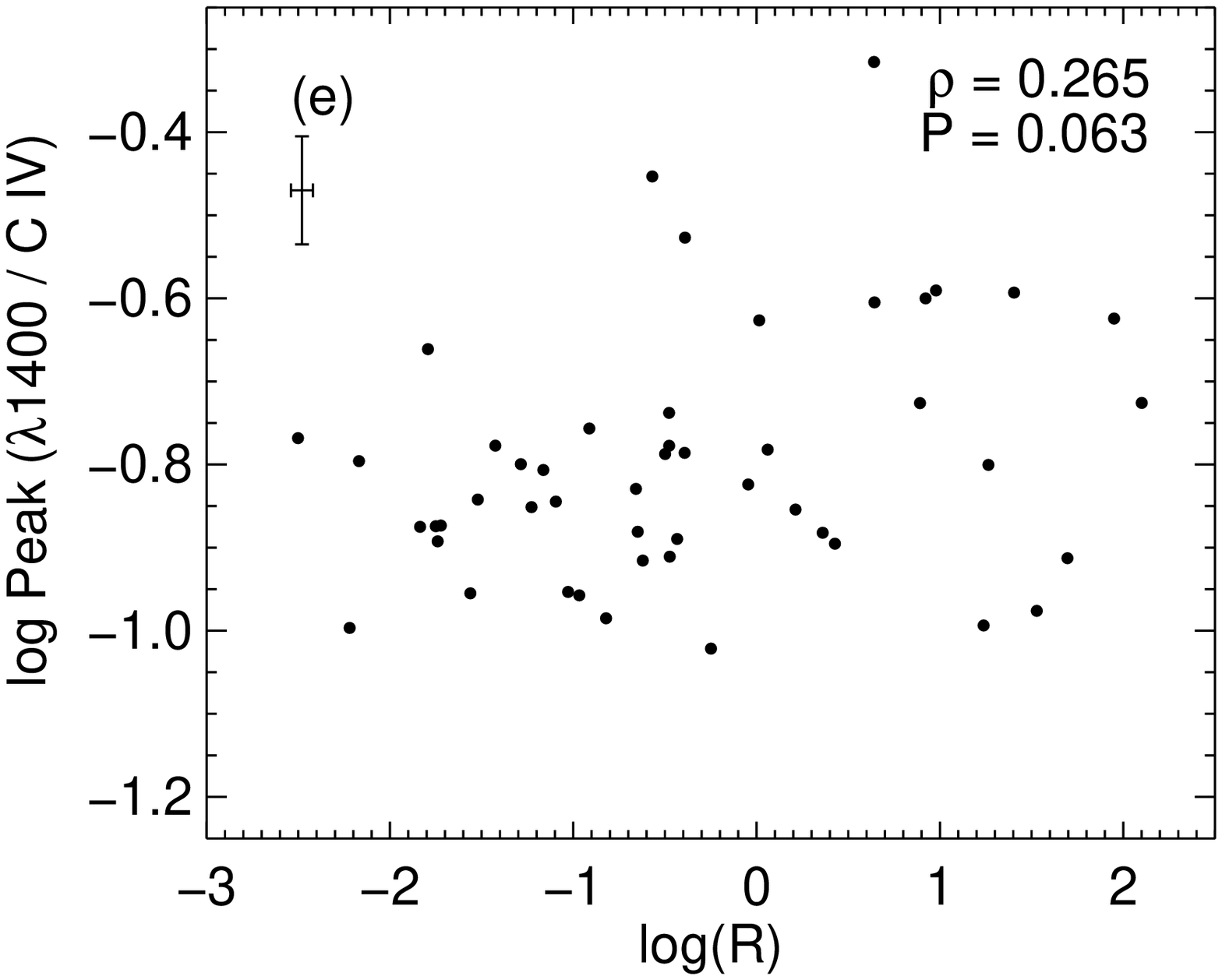}
\end{minipage}
\hspace{0.7cm}
\begin{minipage}[!b]{7cm}
\centering
\includegraphics[width=7cm]{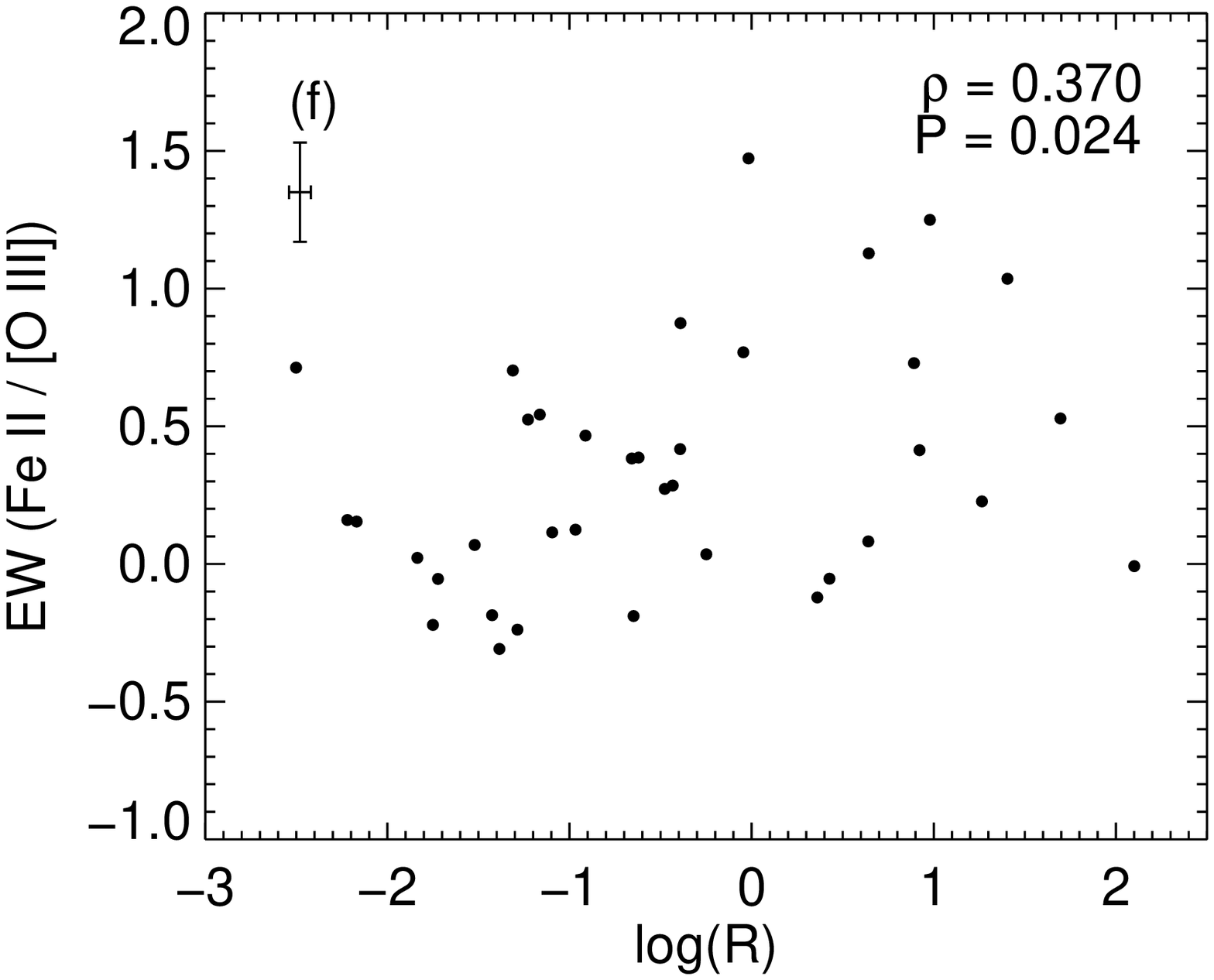}
\end{minipage}
\hspace{0.7cm}
\begin{minipage}[!b]{7cm}
\centering
\includegraphics[width=7cm]{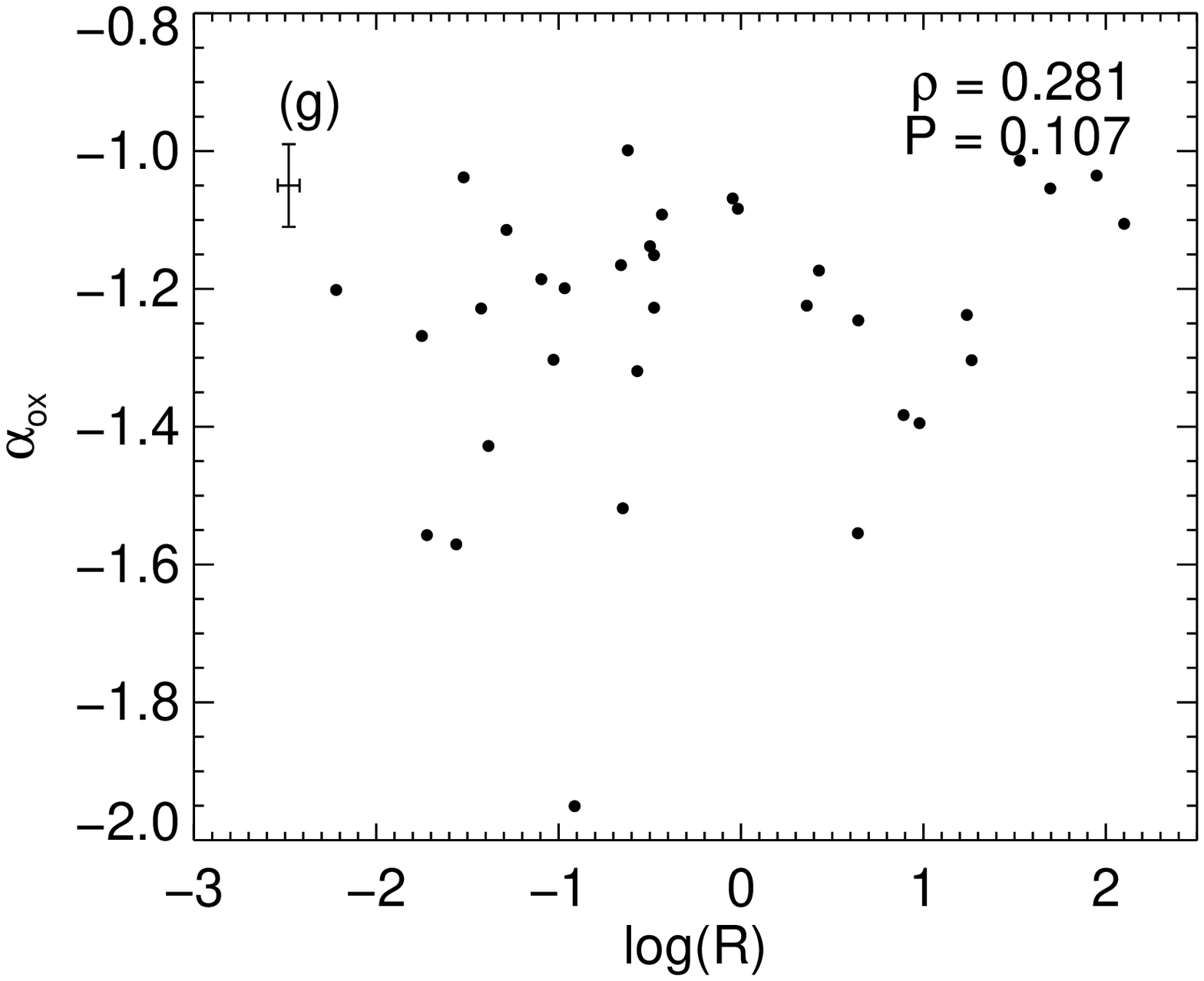}
\end{minipage}     
\hspace{0.7cm}       
\begin{minipage}[!b]{7cm}
\centering
\includegraphics[width=7cm]{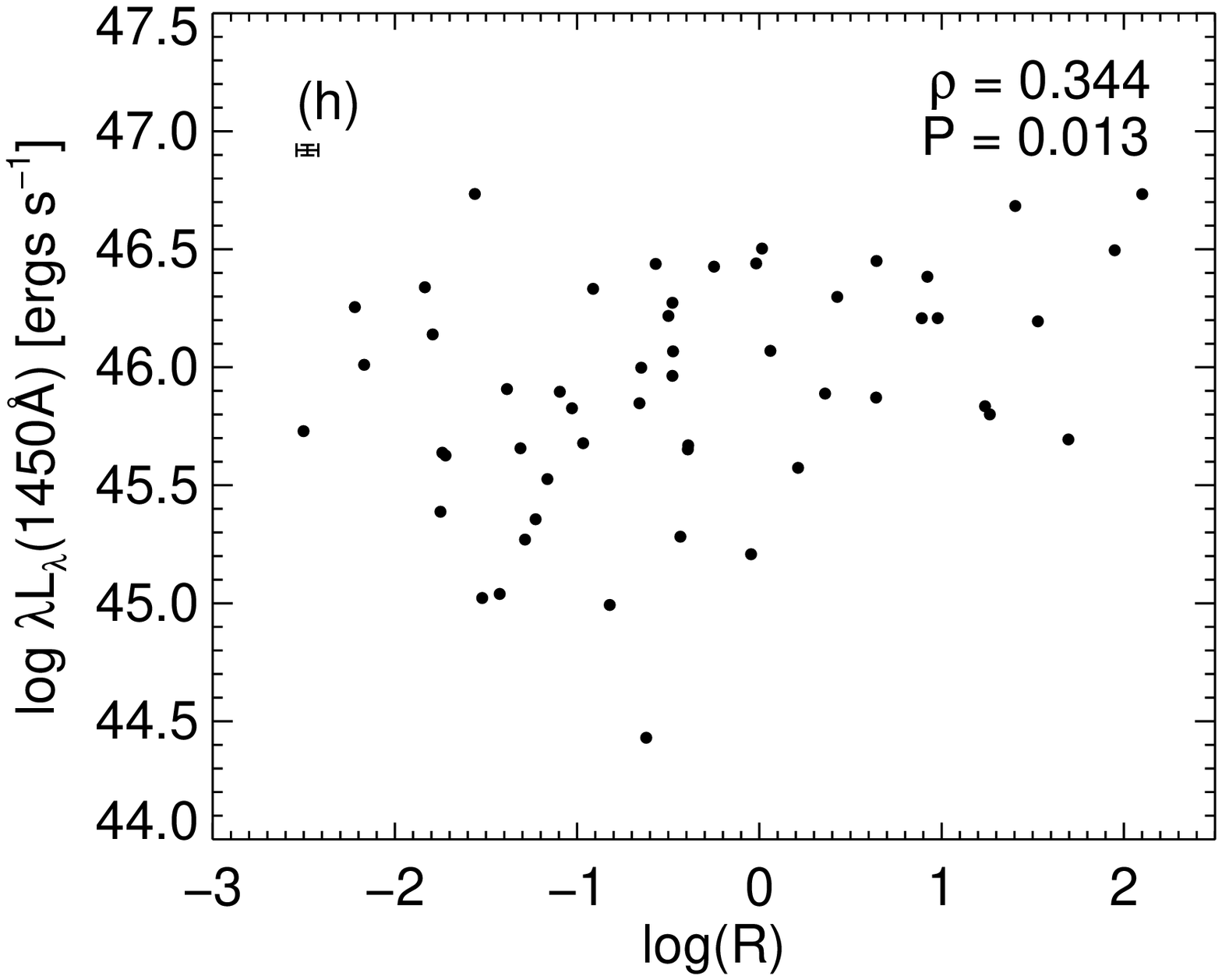}
\end{minipage}             
\caption{Correlations including selected spectral measurements and the orientation indicator  log$\,R$ illustrate the statistics in Table~\ref{tab:corr}.  More edge-on sources are located on the left side of each figure.  The Spearman Rank correlation coefficient and associated probability of finding these distributions of points by chance are listed in the upper right corner of each plot.  Two EV1 indicators, the FWQM of \CIV, and the monochromatic luminosity show marginally significant orientation dependencies in panels (d), (f), (c), and (h), respectively.  The shape and blueshift of \CIV\ and $\alpha_{ox}$ in panels (a), (b), and (g) are notable for their lack of orientation dependence.}
\label{fig:corr}
\end{figure*}

\begin{figure*}
\begin{minipage}[!b]{8.cm}
\centering
\includegraphics[width=8.9cm]{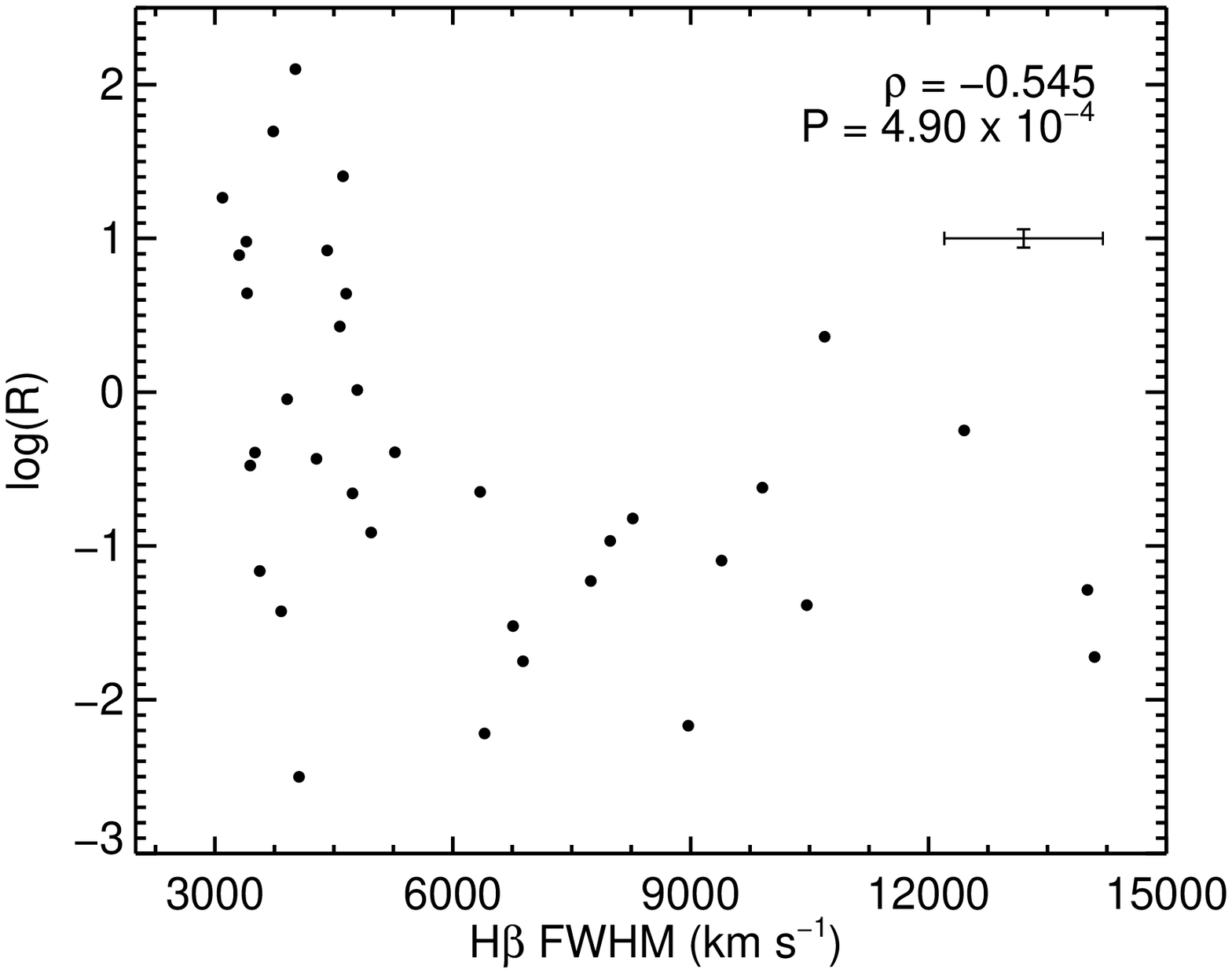}
\end{minipage}
\hspace{0.4cm}
\begin{minipage}[!b]{8.cm}
\centering
\includegraphics[width=8.9cm]{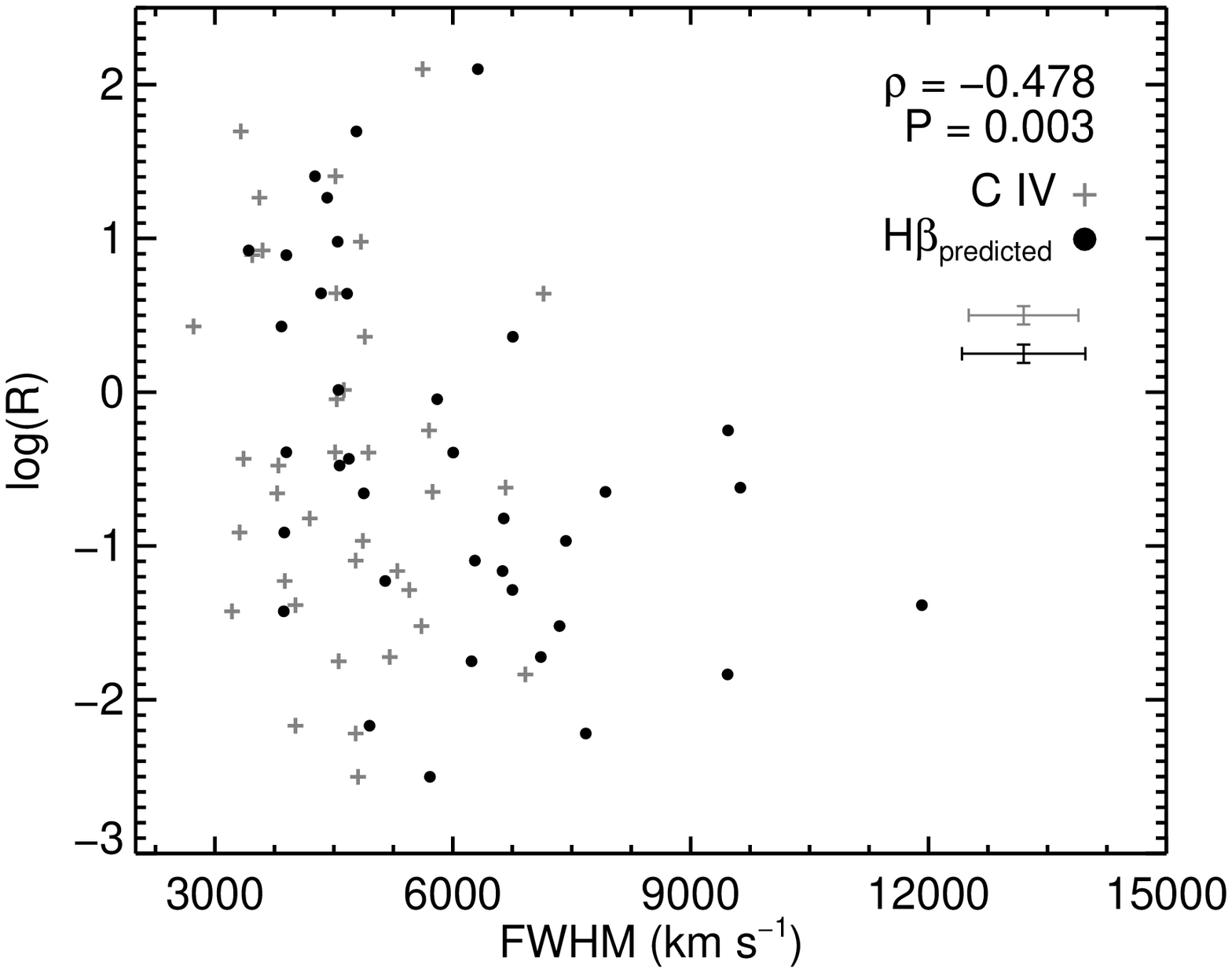}
\end{minipage}
\caption{Radio core dominance, log~$R$, versus FWHM.  The left panel shows the orientation dependence observed for \Hb\ FWHM.  The right panel shows the ``before'' and ``after'' orientation dependence for the \CIV\ and predicted \Hb.  Gray crosses indicate the original \CIV\ values and solid black circles indicate the values predicted for \Hb\ based on measurements of \CIV\ FWHM and the ratio Peak($\lambda$1400/\CIV).  More edge-on sources are located to the bottom in this figure.  Before, \CIV\ FWHM has no significant dependence on orientation and afterwards the dependence is significant at approximately the 3$\sigma$ level and similar to what is observed for \Hb.}
\label{fig:orient}
\end{figure*}

In order to better appreciate the orientation dependencies of the many spectral measurements often made for \CIV, we created the Spearman Rank (RS) correlation matrix displayed in Table~\ref{tab:corr}.  We illustrate the correlations with orientation, highlighted in bold in the table, in Figures~\ref{fig:corr} and \ref{fig:orient}.  Not all objects in the sample have measurements of every parameter, so the number of sources used in each correlation is indicated in the table.  In both the table and figures we display the RS statistic ($\rho$) and the probability ($P$) that the observed distribution of points will be found by chance.  We define a significant correlation as one having $P<0.01$ and a marginally significant correlation as one have a $0.01<P<0.05$.    

We find a significant correlation ($P=0.003$) between the FWHM predicted for \Hb\ based on the FWHM of \CIV\ and the ratio Peak($\lambda$1400/\CIV).  In addition to correlations between the EV1 indicators (the EW ratio of \FeII\ to \OIII, the ratio Peak($\lambda$1400/\CIV), $\alpha_{ox}$, and the blueshift, EW, and shape of \CIV) there are several relationships worth noting in Table~\ref{tab:corr}.  In particular, we focus on orientation dependencies as the strength of this sample.  Some EV1 indicators, including the ratio EW(\FeII/\OIII) and the \CIV\ EW, have a marginally significant orientation dependence ($P=0.024$ and 0.041, respectively), while others do not.  This is consistent with previous work in which the optical \FeII\ emission has an orientation dependence \citep{joly91} and the \OIII\ emission may not be completely isotropic \citep{baker95}.  In the case of the \CIV\ EW, the orientation dependence may be introduced by the continuum.  We see no significant orientation dependence in either the blueshift ($P=0.470$) or the shape of the \CIV\ line ($P=0.516$), where either might have been expected based on interpretations in previous work \citep[e.g.,][]{richards02,denney12}.  The FWQM of \CIV\ has a marginally significant orientation dependence ($P=0.036$), less significant than was found for the FW20M in \citet{vestergaard00}, but perhaps stronger than might have been anticipated based on the composites of \citet{fine11}.  We are optimized to study the orientation dependencies of measured parameters in this sample and we caution against reading too much into the correlations between different EV1 indicators.  They are interesting and worth noting, but we encourage the reader to keep in mind that, as all objects in this sample are RL, they tend toward one end of EV1 and not representative of the full range possible for these properties.

To illustrate the results of the correlation analysis, each parameter is shown versus radio core dominance in Figure~\ref{fig:corr}.  In this figure, the more edge-on sources are located to the left.  The monochromatic luminosity at 1450 \AA\ shows an orientation dependence that is on the verge of being considered significant by our criteria.  The FWQM of \CIV\ and two EV1 indicators, EW(\FeII/\OIII) and the \CIV\ EW, show marginally significant orientation dependencies in panels (c), (d), and (f), respectively.  The notable lack of an orientation dependences that is observed for the \CIV\ shape and blueshift is shown in panels (a) and (b).  The lack of an orientation dependence noted for $\alpha_{ox}$ is illustrated in panel (g).



From Equation~\ref{eqn:EV1}, which is copied directly from Paper I, we calculated the FWHM predicted for \Hb\ based on \CIV\ and the ratio Peak($\lambda$1400/\CIV).  This quantity is shown by those authors to be a better measurement of the characteristic velocity of the virialized gas in the BLR than the FWHM of \CIV\ alone based on its improved agreement with the FWHM of \Hb.  In Figure~\ref{fig:orient} we show the change in orientation dependence from the FWHM of \CIV\ to that of the predicted \Hb.  As expected \citep[e.g.,][]{vestergaard00}, the FWHM of \CIV\ has no orientation dependence.  The FWHM predicted for \Hb\ using the \CIV\ FWHM and Peak($\lambda$1400/\CIV) depends on orientation in a way that appears consistent with the orientation dependence observed for the \Hb\ line width \citep[e.g.,][]{wills86,runnoe13a}.  The significant correlation persists with $P=0.004$ when the point at $\sim12,000$~km~s$^{-1}$, which may appear to drive the correlation, is removed.

\subsection{Composite spectra}

\begin{figure*}
\begin{center}
\includegraphics[width=17.5 truecm]{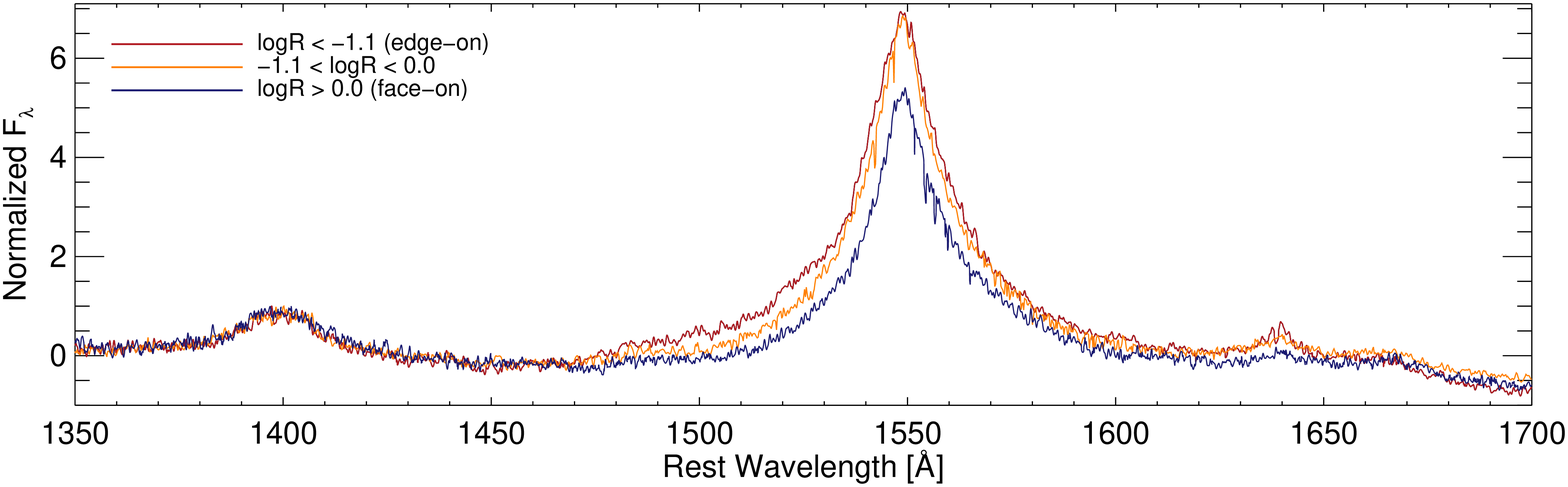}
\end{center}
\caption{Composites of the 1400 \AA\ feature and the \CIV\ line, binned by orientation.  There are 17 objects in the face-on (blue) and edge-on (red) composites and 18 in the intermediate (orange) composite.  The edge-on composite is notably brighter than the face-on composite and modestly broader as well.}
\label{fig:composite}
\end{figure*}

Composite spectra provide another way of investigating the orientation dependencies of different spectral properties.  Because this sample is built to study orientation, we create three composites: an ``edge-on'' composite, a ``face-on'' composite, and an intermediate composite.  To do this, we bin objects by radio core dominance, so in reality for these Type 1 objects, the edge-on composite is not completely edge-on (90$^{\circ}$) because this view is blocked by the dusty torus.  There are 17 objects in the edge-on composite (log$\,R<-1.1$), 18 in the intermediate composite ($-1.1<\,$log$\,R<0.0$), and 17 objects in the face-on composite (log$\,R>0$).

The composites are created following the method of \citet{dipompeo12b}.  We normalize each spectrum by its average flux in the continuum window $1425-1475$ \AA\ to remove luminosity effects, then the spectra in each bin are averaged together with 3$\sigma$ clipping to create the composite spectrum.  The inclusion of 3$\sigma$ clipping can eliminate spikes associated with noise or other artificial features in individual spectra, but odes not significantly change the composites or results that we draw from them in this case.  We also computed median composites, but do not present them here as they provide no additional information.  The composites have been normalized to the peak of the 1400 \AA\ feature in Figure~\ref{fig:composite}, which shows little variation between objects \citep[e.g.,][]{wills93,richards02}, an important distinction because other normalizations (e.g., continuum subtraction or normalizing to the \CIV\ peak) can obscure the differences we hope to observe.  

The correlations between UV spectral properties and orientation that were presented in Table~\ref{tab:corr} can also be seen in the composite spectra shown in Figure~\ref{fig:composite}.  The edge-on and face-on composites clearly have different ratios of the 1400 \AA\ peak to the peak of \CIV, though there is no clear shape or blueshift difference between them.  Finally, the marginal dependence on FWQM of \CIV\ on orientation can perhaps be seen in the variation of the blue wing of the \CIV\ line that persists even when the \CIV\ peaks are normalized to unity.      

The edge-on and intermediate composites are more similar than the intermediate and face-on composites.  This may be due to the fact that the edge-on composite does not include any truly edge-on sources because of the nature of Type 1 objects.  Thus, the edge-on composite is composed of sources that are more similar to those in the intermediate composite.
 
\section{Results and discussion}
\label{sec:discussion}
\subsection{Results}
In quasars, the characteristics of the UV spectra are determined by multiple parameters, including the mass of the central black hole, luminosity, orientation, and the unknown driver of EV1.  These things are not necessarily entirely independent and their effects are difficult to separate.   This sample is optimized to address one of these parameters, orientation in RL quasars, but a full understanding of all these parameters and the relationships between them is necessary to further our ability to make accurate and precise estimates of fundamental properties in quasars. 

The significant orientation dependence of the FWHM predicted for \Hb\ based on the FWHM of \CIV\ and the ratio Peak($\lambda$1400/\CIV) is one highlight of this analysis.  It has not been possible to separate the virial \CIV\ emission from a non-reverberating, low-velocity contaminating component in single-epoch spectra.  This analysis suggests that there is orientation information present in UV spectra that we can begin to tease out following the prescription presented in Paper I.  Once measuring the velocity of the virialized \Hb\ gas from UV spectra comes into standard practice, it will also be important to develop an orientation correction to \CIV-based black hole masses in RL sources similar to the \Hb\ correction presented in \citet{runnoe13a}. 

The orientation dependencies that we find in this sample suggest that, while orientation is not the primary driver of EV1, EV1 may not be completely independent of orientation either.  Our analysis suggests that orientation plays an important role and has an affect on some parameters that indicate EV1, most interestingly for this work EW(\FeII/\OIII) \citep[e.g.,][]{joly91,baker95}, even if it is not solely responsible for an object's location along EV1.  This does not mean that orientation is the physical driver of EV1.  In fact, it is clear from panel (f) of Figure~\ref{fig:corr} that a range in EW(\FeII/\OIII) is possible at fixed orientation.  Sample size and selection may play a role in the apparent disagreement with investigations that find EV1 is completely independent of orientation \citep[e.g.,][]{boroson92}. 

One caveat of this work is related to our sample of exclusively RL quasars, which tend to fall on one end of EV1.  At this time, orientation can only be estimated in RL sources, hence our sample selection, but it does not follow that the results may necessarily be extrapolated to apply to RQ quasars.  The sample is limited in the range of EV1 parameters, like \CIV\ blueshift \citep[e.g.,][]{marziani96,sulentic00a,sulentic07}.  Furthermore, there have been indications that EV1 indicators, like the blueshift of CIV, are determined by the accretion engine in RQ objects but may have an additional source in RL objects \citep[e.g.,][]{punsly10,punsly13}, so we advise appropriate caution when considering these results.

\subsection{Comparison to previous work}
\subsubsection{Correlations}
In this sample we do not observe an orientation dependence in the blueshifts of the \CIV\ line.  An orientation dependence was predicted by the model of \citet{richards02}, where the blueshift would arise from an outflowing wind in the plane of the disk viewed optimally in edge-on objects.  Our analysis is inconsistent with this picture.  \citet{richards11} presents a new version of the disk-wind model where the relationship between the disk and wind components are motivated by changes in the SED, as measured by $\alpha_{ox}$.  In these cases, the relative strength of the X-ray emission compared to the optical/UV determines whether the disk or wind components dominates the source.  In a wind-dominated object, the relatively weak X-rays permit a strong wind to form whereas in a disk-dominated object the relatively strong X-rays inhibit the line-driven wind.  In this new version of the disk-wind model no orientation dependence is observed by \citet{richards11} for the blueshift or EW of \CIV, which is more consistent with what we observe.  In fact, the marginally significant orientation dependence ($P=0.041$) that we see in the EW of \CIV\ may rather be due to the orientation dependence of the continuum emission noted by \citet{runnoe13a} and shown in Figure~\ref{fig:corr}.  Though our sample selection is not optimized to address correlations between the primary indicators for the \citet{richards11} model, they are worth noting.  We find a significant correlation between the EW of \CIV\ and $\alpha_{ox}$ ($P=0.01$), but none between the blueshift of \CIV\ and $\alpha_{ox}$ ($P=0.395$).  The latter may be because our RL quasars have a small range in \CIV\ blueshift, typically $\pm500$~km~s$^{-1}$ compared to $\pm1000$~km~s$^{-1}$ found in \citet{richards11}.
	
\begin{figure}
\begin{center}
\includegraphics[width=8.9 truecm]{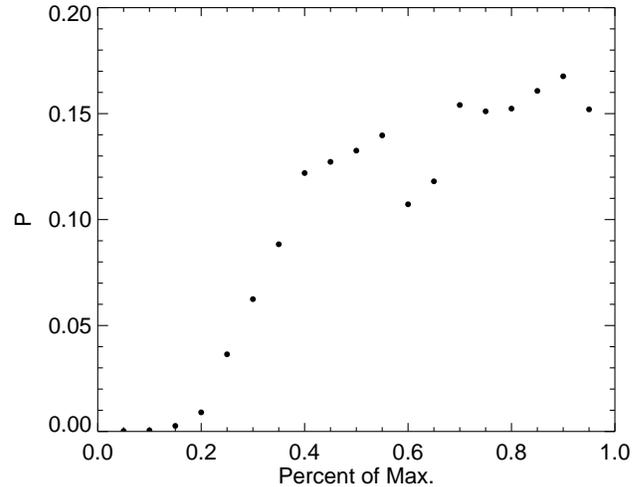}
\end{center}
\caption{The probability that the observed distribution of full-width at percent-maximum versus radio core dominance was found by chance versus the percentage of the maximum flux at which the line width was measured.  This suggests that line widths measured closer to the base of the emission line are dominated by the virial component of the line rather than the low-velocity contaminating component.}
\label{fig:fwpm}
\end{figure}

We find that the orientation dependence of the FWQM of \CIV\ is less significant than the dependence found for the FW20M by \citet{vestergaard00}.  This motivates the question of how the orientation dependence of the velocity line width depends on where in the line (in terms of the percentage of maximum) the width is measured.  Figure~\ref{fig:fwpm} addresses this question and shows that a line width measured closer to the base of the \CIV\ emission line will have a more significant orientation dependence.  This can be interpreted as evidence that as you move towards the base of the \CIV\ emission line, the virial component dominates over the low-velocity contaminating component.  Practically speaking, measuring line widths near the base of the emission line can be difficult due to issues associated with separating the continuum and line emission in the wings of the line.

The lack of an observed orientation dependence in the shape of the \CIV\ line, where we have measured $S=\textrm{FWHM}/\sigma_{l}$, could have significant implications for the \citet{richards11} version of the disk-wind model.  The picture of \citet{richards11} suggests that wind-dominated sources have weak, boxy \CIV\ profiles because these profiles typically have the largest blueshifts.  However, \citet{denney12} finds that it is the emission that contributes to the weak boxy profiles that reverberates in response to changes in continuum emission, behavior expected of the disk component.  Thus, the disk-wind model is still applicable, but only if the strong, peaky \CIV\ profiles are due to large contamination from a wind.  In this case, there is another issue.  The wind model preferred by \citet{richards11} is that of \citet{murray95}.  In such a picture, the wind is expected to be launched from the high-velocity, interior region of the disk.  However, the gas that emits contaminating emission from the wind that creates the peaky profile is low-velocity.  In order to reconcile this model with observations, it is necessary for the wind to have a line-of-sight orientation dependence leading \citet{denney12} to suggest that the shape of the \CIV\ line should have an orientation dependence.  We do not observe such a dependence for the shape of the \CIV\ line, indicating an inconsistency either with the disk-wind model or the expected orientation dependence.  By varying parameters in the disk-wind model, \citet{murray97} are able to reproduce the \CIV\ composite spectra of \citet{wills93} including a profile with a narrow, peaky contaminating core.  It is also true that our sample tends toward one end of EV1-space as it contains only RL objects.  It is possible that this prevents us from uncovering an orientation dependence in the \CIV\ shape because RL objects present a limited range of profile shapes \citep[e.g.,][]{sulentic00a}.

We further consider the implications of the combined observations of \citet{richards11} and \citet{denney12}.  The \citet{richards11} model where weak, boxy \CIV\ profiles with large blueshifts are indicative of a strong wind component is inconsistent with the observation in \citet{denney12} that the reverberating emission has a weak, boxy profile.  A disk-wind model may still be employed if the wind-dominated sources are rather those with strong, peaky \CIV\ profiles.  This raises the following question: why would emission from the BLR that reverberates in response to the continuum show the strongest blueshifts?  Similarly, why would the wind component be seen near the systemic redshift of the source?  Investigations into these questions include, recently, \citet{gaskell13}, who suggest that the entire BLR flows inward and electron and Rayleigh scattering off of the BLR and the torus are responsible for the observed blueshifts and consistent with reverberation mapping results.

\subsubsection{Composites}
Orientation composites of the \CIV\ region have been presented recently by \citet{fine11} and \citet{dipompeo12b}.  In both cases, the composites were binned by radio spectral index, an orientation indicator similar to radio core dominance.    Radio spectral index ($\alpha_{r}$) indicates whether emission from the flat-spectrum core or steep-spectrum lobes is dominating emission at radio frequencies.  Unlike measurements of log$\,R$, measuring $\alpha_{r}$ does not require that the object be spatially resolved and so may be easier to measure in large numbers of objects.  The edge-on and face-on composites of \citet{dipompeo12b}, which were calculated via a method nearly identical to ours, look very similar to those presented here.  \citet{fine11} normalize their composites to the peak of the \CIV\ line, thus erasing any difference in the strength of the \CIV\ emission, but the lack of a shape dependence on orientation can still be seen there.  In fact, their edge-on and face-on composites are so similar that they call into question the orientation dependence found by \citet{vestergaard00}.  

We find that the FWQM orientation dependence is subtle, which may be the reason that \citet{fine11} were unable to recover it.  It is possible that creating their composites with only two bins, $\alpha>-0.5$ and $\alpha<-0.5$, may have obscured the effect by putting similar intermediate objects in each bin.  This seems unlikely, however, because when we construct composites binned in the same way and normalize them to the peak of \CIV\ we still see a subtle width difference.  The radio spectral index measurements have the potential to introduce additional scatter and obscure the variation in FWQM with orientation.  In order to measure radio spectral index for large numbers of objects it is necessary to turn to survey data.  As a result, \citet{fine11} measure radio spectral index from only two frequencies that were not taken simultaneously, as opposed to at least three points or simultaneous measurements as in \citet{dipompeo12b}.  Most likely, however, is that their sample selection is not optimized to study orientation as ours is and may include other effects, for example, variation in the intrinsic luminosity of the source, that swamp out the orientation dependence.

Another point of interest in the \CIV\ composites is the fact that the face-on composite has a weaker \CIV\ line with a smaller EW, peak flux, and flux in the emission line than the edge-on composite.  The EW effect could potentially be due to an orientation dependence in the continuum luminosity, but the fact that the behavior persists in continuum subtracted and continuum normalized versions of the composite make it more likely that this is a real effect.  \citet{fine11} see this effect in some oxygen narrow lines and suggest a configuration in which the line-emitting gas is confined to clouds.  In this case, the majority of the line emission comes from the illuminated face oriented towards the continuum source rather than the optically thick, cool outer face.  \citet{dipompeo12b} suggests that the weaker \CIV\ line in their face-on composite, similar to what we observe in our composites, indicates that this effect may be present for the \CIV\ emitting gas as well.  In that case, \CIV\ emission is weak in the face-on composite because our view is directed at the cool outer cloud faces, whereas in the edge-on composite we had a less obstructed view of the illuminated cloud faces.  We note that it is not necessary to confine this explanation to models with discrete clouds, as it is rather the illumination effect that is relevant for our observations.  

\section{Conclusions}
\label{sec:conclusion}
We investigate spectral properties commonly measured in the \CIV\ spectral region in a sample specifically built for studying orientation.  The orientation subsample is a subset of the \citet{shang11} SED atlas and includes quasi-simultaneous optical/UV spectrophotometry and measurements of radio core dominance, an orientation indicator that estimates the amount of beaming in the radio core.  We perform a correlation analysis and create composite spectra for edge-on and face-on orientations in order to unravel the orientation dependencies of the \CIV\ line and spectral measurements made at nearby wavelengths.  Our main results are as follows:

\begin{itemize}

\item We find that the FWHM predicted for \Hb\ in \citet{runnoe13c} depends on orientation significant at the 3$\sigma$ level.  This result demonstrates that the behavior of the predicted \Hb\ velocity width is consistent the observed line and indicates that some orientation information can be extracted from UV spectra.

\item The FWQM has a marginally significant orientation dependence, less than was found for the FW20M by \citet{vestergaard00} but more than might be expected based on the \CIV\ composites of \citet{fine11}.

\item The blueshift of \CIV\ does not depend on orientation.  The model of \citet{richards02}, which predicts that the blueshift is created by an equatorial outflowing wind and predicts that edge-on objects should have the largest blueshifts, is inconsistent with these observations.  The version of the disk-wind model presented by \citet{richards11} is more consistent with observations as the EW of \CIV\ has only a marginally significant orientation dependence that may be the result of the continuum emission depending on orientation.

\item The shape of \CIV, where $S=\textrm{FWHM}/\sigma_{l}$, does not depend on orientation in this sample.  \citet{denney12} predicts an orientation dependence for the shape \CIV\ line in order to reconcile observations with the model of \citet{richards11}.  It is possible that the orientation dependence exists, but we are unable to uncover it due to a limited range in \CIV\ shapes that results from our RL sample tending towards one end of EV1.

\item We find a marginally significant orientation dependence the optical EV1 indicator EW(\FeII/\OIII).  This suggests that, though an object's location in EV1 is not determined based only on its orientation, orientation does affect some parameters that are included in EV1.

\end{itemize}

\section*{Acknowledgments}
We wish to thank the anonymous referee and Z. Shang acknowledges support by National Basic Research Program of China (973 Program 2013CB834902), Tianjin Distinguished Professor Funds and the National Natural Science Foundation of China (Grant No. 10773006).

\bibliographystyle{/Users/jrunnoe/Library/texmf/bibtex/bst/mn2e}
\bibliography{./all.101713}
\clearpage

\onecolumn 
{\centering
\begin{landscape}
{\scriptsize
\renewcommand{\thefootnote}{\alph{footnote}}
\begin{ThreePartTable}
\begin{longtable}{lcccrcccccccr}
\caption{Spectral properties\label{tab:measurements}
}\\
Object
 & 
EW(\FeII/\OIII)
 & 
\fwhmhb\
 & 
\CIV\ Shape
 & 
\CIV\ blueshift
 & 
FWQM$_{\rm{C\,{\sc IV}}}$
 & 
\fwhmciv\
 & 
CIV EW
 & 
$\rm{FWHM}_{\rm{H}\beta,\rm{predicted}}$
 & 
Peak($\lambda1400$/\CIV)
 & 
$\alpha_{ox}$
 & 
log$\,\lambda$L$_{\lambda}$(1450 \AA)
 & 
log$\,R$
 \\ 
        & \AA\ & km s$^{-1}$ && km s$^{-1}$ & km s$^{-1}$ & km s$^{-1}$ & \AA\ &km s$^{-1}$ &&& ergs s$^{-1}$ & \\
 (1) & (2) & (3) & (4) & (5) & (6) & (7) & (8) & (9) & (10) & (11) & (12) & (13) \\
\hline
\endfirsthead
\multicolumn{13}{c}
{\tablename\ \thetable\ -- \textit{Continued}} \\
Object
 & 
EW(\FeII/\OIII)
 & 
\fwhmhb\
 & 
\CIV\ Shape
 & 
\CIV\ blueshift
 & 
FWQM$_{\rm{C\,{\sc IV}}}$
 & 
\fwhmciv\
 & 
CIV EW
 & 
$\rm{FWHM}_{\rm{H}\beta,\rm{predicted}}$
 & 
Peak($\lambda1400$/\CIV)
 & 
$\alpha_{ox}$
 & 
log$\,\lambda$L$_{\lambda}$(1450 \AA)
 & 
log$\,R$
 \\ 
        & \AA\ & km s$^{-1}$ && km s$^{-1}$ & km s$^{-1}$ & km s$^{-1}$ & \AA\ &km s$^{-1}$ &&& ergs s$^{-1}$ & \\
\hline
\endhead
\hline
\multicolumn{13}{r}{\textit{Continued on next page}} \\
\endfoot
\hline
\endlastfoot
\hline
3C 110   
	& 
    1.08
 & 
       12450
 & 
    1.08
 & 
$-$307
 & 
       13696
 & 
        5700
 & 
  107.78
 & 
        9471
 & 
    0.10
 & 
\nodata
 & 
   46.43
 & 
$-$0.249
 \\ 
3C 175   
	& 
    1.05
 & 
       20925
 & 
    1.30
 & 
        1072
 & 
       14577
 & 
        6915
 & 
   55.59
 & 
        9466
 & 
    0.13
 & 
\nodata
 & 
   46.34
 & 
$-$1.836
 \\ 
3C 186   
	& 
\nodata
 & 
\nodata
 & 
    1.39
 & 
$-$185
 & 
       12526
 & 
        6290
 & 
   66.50
 & 
        7614
 & 
    0.17
 & 
\nodata
 & 
   46.07
 & 
   0.060
 \\ 
3C 207   
	& 
    2.61
 & 
        3505
 & 
    1.46
 & 
         227
 & 
        8157
 & 
        4935
 & 
   86.32
 & 
        6004
 & 
    0.16
 & 
\nodata
 & 
   45.65
 & 
$-$0.393
 \\ 
3C 215   
	& 
    1.17
 & 
        6760
 & 
    1.11
 & 
$-$51
 & 
       10043
 & 
        5605
 & 
  204.60
 & 
        7346
 & 
    0.14
 & 
$-$1.04
 & 
   45.02
 & 
$-$1.521
 \\ 
3C 232   
	& 
    1.21
 & 
        4655
 & 
    1.76
 & 
         155
 & 
       12291
 & 
        7145
 & 
   33.54
 & 
        4666
 & 
    0.48
 & 
$-$1.55
 & 
   45.87
 & 
   0.641
 \\ 
3C 254   
	& 
    0.88
 & 
       14095
 & 
    1.02
 & 
$-$215
 & 
       13113
 & 
        5205
 & 
  172.43
 & 
        7110
 & 
    0.13
 & 
$-$1.56
 & 
   45.63
 & 
$-$1.722
 \\ 
3C 263   
	& 
    2.93
 & 
        4970
 & 
    0.74
 & 
$-$284
 & 
        6219
 & 
        3310
 & 
   75.51
 & 
        3874
 & 
    0.18
 & 
$-$1.95
 & 
   46.33
 & 
$-$0.912
 \\ 
3C 277.1   
	& 
    0.65
 & 
        3835
 & 
    0.96
 & 
$-$176
 & 
        5783
 & 
        3215
 & 
  106.47
 & 
        3867
 & 
    0.17
 & 
$-$1.23
 & 
   45.04
 & 
$-$1.425
 \\ 
3C 281   
	& 
    1.33
 & 
        7985
 & 
    0.94
 & 
$-$460
 & 
       12980
 & 
        4865
 & 
  119.67
 & 
        7426
 & 
    0.11
 & 
$-$1.20
 & 
   45.68
 & 
$-$0.967
 \\ 
3C 288.1   
	& 
    1.43
 & 
        8970
 & 
    0.84
 & 
$-$272
 & 
        9905
 & 
        4015
 & 
   42.32
 & 
        4949
 & 
    0.16
 & 
\nodata
 & 
   46.01
 & 
$-$2.169
 \\ 
3C 334   
	& 
    0.65
 & 
        6345
 & 
    1.44
 & 
$-$21
 & 
       11404
 & 
        5745
 & 
   74.98
 & 
        7925
 & 
    0.13
 & 
$-$1.52
 & 
   46.00
 & 
$-$0.648
 \\ 
3C 37   
	& 
    1.93
 & 
        4280
 & 
    1.04
 & 
$-$169
 & 
        6934
 & 
        3360
 & 
  252.63
 & 
        4689
 & 
    0.13
 & 
$-$1.09
 & 
   45.28
 & 
$-$0.433
 \\ 
3C 446   
	& 
\nodata
 & 
\nodata
 & 
    0.90
 & 
         143
 & 
        8022
 & 
        3390
 & 
   76.40
 & 
        5304
 & 
    0.11
 & 
$-$1.01
 & 
   46.19
 & 
   1.528
 \\ 
3C 47   
	& 
    0.58
 & 
       14005
 & 
    1.03
 & 
$-$432
 & 
       10431
 & 
        5450
 & 
  172.91
 & 
        6751
 & 
    0.16
 & 
$-$1.11
 & 
   45.27
 & 
$-$1.286
 \\ 
4C 01.04   
	& 
    2.43
 & 
        9905
 & 
    1.46
 & 
           1
 & 
       12443
 & 
        6665
 & 
  270.36
 & 
        9626
 & 
    0.12
 & 
$-$1.00
 & 
   44.43
 & 
$-$0.621
 \\ 
4C 06.69   
	& 
    0.98
 & 
        4015
 & 
    1.31
 & 
$-$1002
 & 
       11478
 & 
        5620
 & 
   45.16
 & 
        6315
 & 
    0.19
 & 
$-$1.11
 & 
   46.73
 & 
   2.101
 \\ 
4C 10.06   
	& 
    2.42
 & 
        4735
 & 
    0.83
 & 
$-$272
 & 
        9196
 & 
        3785
 & 
  114.57
 & 
        4877
 & 
    0.15
 & 
$-$1.17
 & 
   45.85
 & 
$-$0.658
 \\ 
4C 12.40   
	& 
    3.49
 & 
        3565
 & 
    1.18
 & 
$-$7
 & 
       10036
 & 
        5300
 & 
   97.16
 & 
        6627
 & 
    0.16
 & 
\nodata
 & 
   45.53
 & 
$-$1.163
 \\ 
4C 19.44   
	& 
    0.88
 & 
        4575
 & 
    0.75
 & 
$-$147
 & 
        6763
 & 
        2730
 & 
   96.37
 & 
        3838
 & 
    0.13
 & 
$-$1.17
 & 
   46.30
 & 
   0.427
 \\ 
4C 20.24   
	& 
\nodata
 & 
\nodata
 & 
    0.96
 & 
$-$19
 & 
        7050
 & 
        3525
 & 
  185.25
 & 
        4297
 & 
    0.16
 & 
$-$1.14
 & 
   46.22
 & 
$-$0.499
 \\ 
4C 22.26   
	& 
\nodata
 & 
\nodata
 & 
    1.14
 & 
          97
 & 
       10979
 & 
        5015
 & 
  180.90
 & 
        7613
 & 
    0.11
 & 
$-$1.30
 & 
   45.83
 & 
$-$1.028
 \\ 
4C 30.25   
	& 
\nodata
 & 
\nodata
 & 
    1.00
 & 
$-$282
 & 
        7871
 & 
        3730
 & 
  146.34
 & 
        5225
 & 
    0.13
 & 
\nodata
 & 
   45.64
 & 
$-$1.740
 \\ 
4C 31.63   
	& 
   17.77
 & 
        3395
 & 
    1.12
 & 
$-$256
 & 
       10734
 & 
        4840
 & 
   39.60
 & 
        4547
 & 
    0.26
 & 
$-$1.39
 & 
   46.21
 & 
   0.979
 \\ 
4C 39.25   
	& 
    1.44
 & 
        6400
 & 
    1.08
 & 
         772
 & 
       10352
 & 
        4775
 & 
   79.99
 & 
        7677
 & 
    0.10
 & 
$-$1.20
 & 
   46.25
 & 
$-$2.220
 \\ 
4C 40.24   
	& 
\nodata
 & 
\nodata
 & 
    1.51
 & 
$-$239
 & 
        7534
 & 
        4920
 & 
  131.38
 & 
        5617
 & 
    0.18
 & 
$-$1.15
 & 
   45.96
 & 
$-$0.477
 \\ 
4C 41.21   
	& 
    1.87
 & 
        3445
 & 
    1.02
 & 
$-$195
 & 
        7874
 & 
        3800
 & 
   96.04
 & 
        4571
 & 
    0.17
 & 
$-$1.23
 & 
   46.27
 & 
$-$0.477
 \\ 
4C 49.22   
	& 
    5.87
 & 
        3910
 & 
    1.31
 & 
          63
 & 
        8357
 & 
        4535
 & 
  178.77
 & 
        5803
 & 
    0.15
 & 
$-$1.07
 & 
   45.21
 & 
$-$0.046
 \\ 
4C 55.17   
	& 
\nodata
 & 
\nodata
 & 
    1.70
 & 
         630
 & 
       11004
 & 
        6420
 & 
   34.57
 & 
       10282
 & 
    0.10
 & 
$-$1.24
 & 
   45.83
 & 
   1.238
 \\ 
4C 58.29   
	& 
\nodata
 & 
\nodata
 & 
    1.36
 & 
          98
 & 
       11837
 & 
        5745
 & 
   39.43
 & 
        8741
 & 
    0.11
 & 
$-$1.57
 & 
   46.73
 & 
$-$1.561
 \\ 
4C 64.15   
	& 
\nodata
 & 
\nodata
 & 
    1.39
 & 
$-$351
 & 
       13256
 & 
        7245
 & 
   68.54
 & 
        7474
 & 
    0.22
 & 
\nodata
 & 
   46.14
 & 
$-$1.793
 \\ 
4C 73.18   
	& 
    1.69
 & 
        3095
 & 
    0.99
 & 
$-$52
 & 
        7946
 & 
        3560
 & 
  111.79
 & 
        4415
 & 
    0.16
 & 
$-$1.30
 & 
   45.80
 & 
   1.265
 \\ 
B2 0742+31   
	& 
    0.76
 & 
       10690
 & 
    1.09
 & 
$-$852
 & 
       11404
 & 
        4890
 & 
  127.78
 & 
        6757
 & 
    0.13
 & 
$-$1.22
 & 
   45.89
 & 
   0.360
 \\ 
B2 1351+31   
	& 
\nodata
 & 
\nodata
 & 
    0.78
 & 
$-$23
 & 
       11931
 & 
        3690
 & 
   47.73
 & 
        5296
 & 
    0.12
 & 
\nodata
 & 
   46.07
 & 
$-$0.474
 \\ 
B2 1555+33   
	& 
\nodata
 & 
\nodata
 & 
    1.12
 & 
$-$259
 & 
        9107
 & 
        4240
 & 
  108.54
 & 
        5646
 & 
    0.14
 & 
\nodata
 & 
   45.57
 & 
   0.212
 \\ 
B2 1611+34   
	& 
\nodata
 & 
        4795
 & 
    1.21
 & 
         765
 & 
        9516
 & 
        4625
 & 
   50.65
 & 
        4557
 & 
    0.24
 & 
\nodata
 & 
   46.50
 & 
   0.014
 \\ 
MC2 0042+101   
	& 
\nodata
 & 
        8270
 & 
    0.91
 & 
$-$184
 & 
       10889
 & 
        4195
 & 
  218.94
 & 
        6642
 & 
    0.10
 & 
\nodata
 & 
   44.99
 & 
$-$0.821
 \\ 
MC2 1146+111   
	& 
    5.04
 & 
        7835
 & 
    0.95
 & 
$-$564
 & 
        8353
 & 
        3715
 & 
   46.76
 & 
\nodata
 & 
\nodata
 & 
\nodata
 & 
   45.66
 & 
$-$1.311
 \\ 
OS 562   
	& 
    5.36
 & 
        3305
 & 
    0.82
 & 
          16
 & 
        6744
 & 
        3470
 & 
   46.38
 & 
        3900
 & 
    0.19
 & 
$-$1.38
 & 
   46.21
 & 
   0.891
 \\ 
PG 1100+772   
	& 
    1.30
 & 
        9390
 & 
    1.00
 & 
         143
 & 
       12663
 & 
        4775
 & 
   79.68
 & 
        6278
 & 
    0.14
 & 
$-$1.19
 & 
   45.90
 & 
$-$1.095
 \\ 
PG 1103-006   
	& 
    7.49
 & 
        5270
 & 
    1.07
 & 
         146
 & 
        8628
 & 
        4515
 & 
   55.29
 & 
        3900
 & 
    0.30
 & 
\nodata
 & 
   45.67
 & 
$-$0.391
 \\ 
PG 1226+023   
	& 
   13.43
 & 
        3405
 & 
    1.22
 & 
$-$468
 & 
        8587
 & 
        4530
 & 
   32.40
 & 
        4338
 & 
    0.25
 & 
$-$1.25
 & 
   46.45
 & 
   0.643
 \\ 
PG 1545+210   
	& 
    0.60
 & 
        6885
 & 
    0.98
 & 
         150
 & 
       11405
 & 
        4560
 & 
  181.05
 & 
        6235
 & 
    0.13
 & 
$-$1.27
 & 
   45.39
 & 
$-$1.750
 \\ 
PG 1704+608   
	& 
    0.49
 & 
       10465
 & 
    0.75
 & 
$-$493
 & 
       12764
 & 
        4015
 & 
   61.20
 & 
       11916
 & 
    0.03
 & 
$-$1.43
 & 
   45.91
 & 
$-$1.385
 \\ 
PG 2251+113   
	& 
    5.16
 & 
        4060
 & 
    1.13
 & 
$-$436
 & 
        9366
 & 
        4805
 & 
   96.91
 & 
        5711
 & 
    0.17
 & 
\nodata
 & 
   45.73
 & 
$-$2.501
 \\ 
PKS 0112-017   
	& 
\nodata
 & 
\nodata
 & 
    1.25
 & 
$-$325
 & 
       10721
 & 
        5030
 & 
   30.81
 & 
        3941
 & 
    0.35
 & 
$-$1.32
 & 
   46.44
 & 
$-$0.569
 \\ 
PKS 0403-13   
	& 
    3.38
 & 
        3735
 & 
    0.84
 & 
         307
 & 
        5431
 & 
        3325
 & 
  138.32
 & 
        4784
 & 
    0.12
 & 
$-$1.05
 & 
   45.69
 & 
   1.696
 \\ 
PKS 0859-14   
	& 
   10.86
 & 
        4615
 & 
    1.21
 & 
$-$830
 & 
        8720
 & 
        4520
 & 
   48.89
 & 
        4261
 & 
    0.26
 & 
\nodata
 & 
   46.68
 & 
   1.404
 \\ 
PKS 1127-14   
	& 
\nodata
 & 
\nodata
 & 
    1.06
 & 
         371
 & 
        7339
 & 
        3695
 & 
   25.93
 & 
        3630
 & 
    0.24
 & 
$-$1.04
 & 
   46.50
 & 
   1.950
 \\ 
PKS 1656+053   
	& 
   29.71
 & 
        3510
 & 
\nodata
 & 
$-$164
 & 
\nodata
 & 
\nodata
 & 
\nodata
 & 
\nodata
 & 
\nodata
 & 
$-$1.08
 & 
   46.44
 & 
$-$0.017
 \\ 
PKS 2216-03   
	& 
    2.59
 & 
        4415
 & 
    0.80
 & 
$-$107
 & 
       10156
 & 
        3600
 & 
   59.31
 & 
        3425
 & 
    0.25
 & 
\nodata
 & 
   46.38
 & 
   0.922
 \\ 
TEX 1156+213   
	& 
    3.34
 & 
        7740
 & 
    0.76
 & 
$-$382
 & 
       10997
 & 
        3880
 & 
  116.09
 & 
        5147
 & 
    0.14
 & 
\nodata
 & 
   45.36
 & 
$-$1.228
 \\ 
\hline
\end{longtable}
\begin{tablenotes}
\small
\item Note $-$ EWs are given in the rest frame.
\end{tablenotes}
\end{ThreePartTable}
} 
\end{landscape}
} 

\twocolumn
\clearpage

\onecolumn 

{\centering
\afterpage{
\begin{landscape}
{\scriptsize
\renewcommand{\thefootnote}{\alph{footnote}}
\begin{ThreePartTable}
\begin{longtable}{rrrrrrrrrrrr}
\caption{Correlation matrix RL subsample}
\label{tab:corr}\\
\hline
 & 
EW(\FeII/\OIII)
 & 
\fwhmhb\
 & 
\CIV\ Shape
 & 
\CIV\ blueshift
 & 
FWQM$_{\rm{C\,{\sc IV}}}$
 & 
\fwhmciv\
 & 
CIV EW
 & 
\fwhmhbpred
 & 
Peak($\lambda1400$/\CIV)
 & 
$\alpha_{ox}$
 & 
$\lambda$L$_{\lambda}$(1450 \AA)
 \\ 
\fwhmhb\
 & 
$-$0.575
(37)
 &

 &

 &

 &

 &

 &

 &

 &

 &

 &

\\
 & 
$<$10$^{-3}$
 & 

 & 

 & 

 & 

 & 

 & 

 & 

 & 

 & 

 & 

\\
\CIV\ Shape
 & 
   0.094
(36)
 & 
$-$0.092
(38)
 &

 &

 &

 &

 &

 &

 &

 &

 &

\\
 & 
   0.586
 & 
   0.583
 & 

 & 

 & 

 & 

 & 

 & 

 & 

 & 

 & 

\\
\CIV\ blueshift
 & 
   0.032
(37)
 & 
$-$0.140
(39)
 & 
   0.174
(51)
 &

 &

 &

 &

 &

 &

 &

 &

\\
 & 
   0.853
 & 
   0.394
 & 
   0.221
 & 

 & 

 & 

 & 

 & 

 & 

 & 

 & 

\\
FWQM$_{\rm{C\,{\sc IV}}}$
 & 
$-$0.459
(36)
 & 
   0.702
(38)
 & 
   0.302
(51)
 & 
$-$0.083
(51)
 &

 &

 &

 &

 &

 &

 &

\\
 & 
   0.005
 & 
$<$10$^{-6}$
 & 
   0.031
 & 
   0.561
 & 

 & 

 & 

 & 

 & 

 & 

 & 

\\
\fwhmciv\
 & 
$-$0.272
(36)
 & 
   0.394
(38)
 & 
   0.796
(51)
 & 
   0.041
(51)
 & 
   0.744
(51)
 &

 &

 &

 &

 &

 &

\\
 & 
   0.108
 & 
   0.014
 & 
$<$10$^{-11}$
 & 
   0.774
 & 
$<$10$^{-9}$
 & 

 & 

 & 

 & 

 & 

 & 

\\
CIV EW
 & 
$-$0.283
(36)
 & 
   0.266
(38)
 & 
$-$0.169
(51)
 & 
$-$0.023
(51)
 & 
$-$0.090
(51)
 & 
$-$0.097
(51)
 &

 &

 &

 &

 &

\\
 & 
   0.095
 & 
   0.107
 & 
   0.237
 & 
   0.871
 & 
   0.529
 & 
   0.499
 & 

 & 

 & 

 & 

 & 

\\
\fwhmhbpred
 & 
$-$0.478
(35)
 & 
   0.658
(37)
 & 
   0.297
(37)
 & 
$-$0.023
(37)
 & 
   0.744
(37)
 & 
   0.697
(37)
 & 
   0.383
(37)
 &

 &

 &

 &

\\
 & 
   0.004
 & 
$<$10$^{-5}$
 & 
   0.074
 & 
   0.890
 & 
$<$10$^{-6}$
 & 
$<$10$^{-5}$
 & 
   0.019
 & 

 & 

 & 

 & 

\\
Peak($\lambda1400$/\CIV)
 & 
   0.485
(35)
 & 
$-$0.531
(37)
 & 
   0.219
(50)
 & 
$-$0.142
(50)
 & 
$-$0.280
(50)
 & 
$-$0.012
(50)
 & 
$-$0.447
(50)
 & 
$-$0.692
(37)
 &

 &

 &

\\
 & 
   0.003
 & 
   0.001
 & 
   0.126
 & 
   0.326
 & 
   0.049
 & 
   0.933
 & 
   0.001
 & 
$<$10$^{-5}$
 & 

 & 

 & 

\\
$\alpha_{ox}$
 & 
   0.227
(26)
 & 
   0.015
(26)
 & 
$-$0.009
(33)
 & 
   0.151
(34)
 & 
$-$0.275
(33)
 & 
$-$0.162
(33)
 & 
   0.442
(33)
 & 
   0.211
(25)
 & 
$-$0.132
(33)
 &

 &

\\
 & 
   0.264
 & 
   0.941
 & 
   0.962
 & 
   0.395
 & 
   0.122
 & 
   0.368
 & 
   0.010
 & 
   0.312
 & 
   0.464
 & 

 & 

\\
$\lambda$L$_{\lambda}$(1450 \AA)
 & 
   0.143
(37)
 & 
$-$0.192
(39)
 & 
   0.080
(51)
 & 
$-$0.024
(52)
 & 
   0.036
(51)
 & 
   0.010
(51)
 & 
$-$0.686
(51)
 & 
$-$0.256
(37)
 & 
   0.298
(50)
 & 
$-$0.220
(34)
 &

\\
 & 
   0.399
 & 
   0.242
 & 
   0.576
 & 
   0.867
 & 
   0.801
 & 
   0.947
 & 
$<$10$^{-7}$
 & 
   0.127
 & 
   0.035
 & 
   0.212
 & 

\\
log$\,R$
 & 
   0.370
(37)
 & 
$-$0.554
(39)
 & 
   0.093
(51)
 & 
   0.102
(52)
 & 
$-$0.294
(51)
 & 
$-$0.211
(51)
 & 
$-$0.287
(51)
 & 
$-$0.478
(37)
 & 
   0.265
(50)
 & 
   0.281
(34)
 & 
   0.344
(52)
\\
 & {\bf
   0.024
} & {\bf
$<$10$^{-3}$
} & {\bf
   0.516
} & {\bf
   0.470
} & {\bf
   0.036
} & {\bf
   0.137
} & {\bf
   0.041
} & {\bf
   0.003
} & {\bf
   0.063
} & {\bf
   0.107
} & {\bf
   0.013
}\\
\hline

\end{longtable}
\begin{tablenotes}
\small
\item Note $-$ The Spearman Rank correlation statistic is listed for each parameter pair, with the probability of finding the observed distribution of points by chance listed underneath.  Not all parameters are available for every source, so the number of objects used in the correlation is given in parentheses.
\end{tablenotes}
\end{ThreePartTable}
} 
\end{landscape}
} 
} 

\twocolumn
\clearpage 

\label{lastpage}
\end{document}